\documentclass{emulateapj}
\usepackage{graphicx}
\usepackage{amsmath}
\usepackage{epstopdf}

\newcommand{\be}{\begin{equation}}
\newcommand{\ee}{\end{equation}}
\newcommand{\ba}{\begin{eqnarray}}
\newcommand{\ea}{\end{eqnarray}}

\newcommand{\nn}{\mbox{} \nonumber \\ \mbox{} }

\begin{document} 

\title{Rotation of Giant Stars}
\author{Yevgeni Kissin}
\affil{Department of Astronomy and Astrophysics, University of Toronto, 50 St. George St., Toronto, ON M5S 3H4, Canada}
\author{Christopher Thompson}
\affil{Canadian Institute for Theoretical Astrophysics, 60 St. George St., Toronto, ON M5H 3H8, Canada}

\begin{abstract}
The internal rotation of post-main sequence stars is investigated, in response to the convective
pumping of angular momentum toward the stellar core, combined with a tight magnetic coupling between 
core and envelope.  The spin evolution is calculated using model stars of initial mass 1, 1.5 and $5\,M_\odot$,
taking into account mass loss on the giant branches. 
We also include the deposition of 
orbital angular momentum from a sub-stellar companion, as influenced by tidal drag along with the excitation
of orbital eccentricity by a fluctuating gravitational quadrupole moment.   A range of angular velocity profiles
$\Omega(r)$ is considered in the envelope, extending from solid rotation to constant specific angular
momentum.  We focus on the backreaction of the Coriolis force,
and the threshold for dynamo action in the inner envelope.
Quantitative agreement with measurements of core rotation in subgiants and post-He core flash stars by {\it Kepler} 
is obtained with a two-layer angular velocity profile:  uniform specific angular momentum where the 
Coriolis parameter ${\rm Co} \equiv \Omega \tau_{\rm con} \lesssim 1$ (here $\tau_{\rm con}$ is the convective time); 
and $\Omega(r) \propto r^{-1}$ where ${\rm Co} \gtrsim 1$.  The inner profile is interpreted in terms of a balance 
between the Coriolis force and angular pressure gradients driven by radially extended convective plumes.  
Inward angular momentum pumping reduces the surface rotation of subgiants, and the need 
for a rejuvenated magnetic wind torque.  The co-evolution of internal magnetic fields and rotation is
considered in Kissin \& Thompson, along with the breaking of the rotational coupling between core and envelope due to heavy
mass loss.
\end{abstract}

\section{Introduction}

Measurements of internal oscillation modes of red giant stars by the {\it Kepler} satellite have been used to infer
rapid rotation in the radiative core  -- in some cases, an order of magnitude faster than at the surface
\citep{becketal2012,mossetal2012}.  This evidence has been interpreted by \cite{CantMBCP2014} in terms of a 
decoupling of the rotation of the core from the convective envelope, which is
assumed to rotate as a solid body.

Here we reconsider the role that convection plays in redistributing angular momentum within a star.  So far
only a limited understanding has been developed of the interaction between rotation and convection.
The most convincing theoretical calculations involve direct numerical simulations (e.g. \citealt{brunp2009}).

Red giants offer a promising test case, because (i) the influence of the Coriolis force is reduced 
(although not entirely eliminated) in comparison with main sequence (MS) stars; (ii) the great depth of
the convective envelope reduces the effect of boundary conditions on the angular velocity profile;
and (iii) feedback from magnetic fields on the rotation profile is less important, especially near the tip of the 
red giant and asymptotic giant branches (RGB and AGB), where most of the envelope rotates slowly.
Disentangling the most important processes that control the rotation profile of a 
star is, for these reasons, easier in red giants than in the Sun.

The internal rotation of giant stars also has important implications for the magnetism and spin of their
white dwarf (WD) remnants.  A basic theme of this paper, and the companion
paper \cite[hereafter Paper II]{KissT2015} 
is that one must consider the co-evolution of rotation and magnetic fields if one is to understand either.  
The strong flow of material from 
the hydrogen-rich envelope, through burning shell(s), and into the core implies that one cannot treat 
core and envelope in isolation from each other.  

This observation leads us to discard models of the rotation which neglect the influence 
of large-scale poloidal magnetic fields that were deposited by previous convective activity.   For example,
the Modules for Experiments in Stellar Astrophysics (MESA) code \citep{paxtetal2011} implements a heuristic prescription for angular momentum redistribution
in the radiative parts of a star, which starts with the current-driven (Tayler) instability of a 
predominantly toroidal magnetic field, leading to a poloidal component that is then wound back by 
differential rotation into the toroidal direction \citep{Spru2002}.  

The nearly solid rotation of the Sun's core has been used to infer the presence of a helical magnetic field \citep{GougM1998}. 
The introduction of a persistent twist to the field not only facilitates hydromagnetic stability 
\citep{brais2004}, but also suppresses differential rotation across poloidal magnetic flux surfaces. 
The minimal poloidal field that will transport angular momentum rapidly enough to compensate changes
in the core mass profile on the MS is only $\sim 10^{-6}$ G; but a field $\sim 10^2$-$10^3$ times stronger
is required to compensate a latitude-depenent convective stress that is partly transmitted through
a tachocline layer into the core (Paper II).  The ascent of the early red giant branch is
completed over a Gyr, not a great deal shorter than the present age of the Sun, meaning that a similarly
weak poloidal magnetic flux would enforce nearly solid rotation in the core of a subgiant 
(e.g. \citealt{maeder14}).  

In a $1$-$2M_\odot$ star, the source of this field differs between the early
and later stages of post-MS expansion.  The envelope expands in mass during the first dredge-up phase, 
so that any remnant field in the outer core would have been deposited by a vigorously convective and 
rapidly rotating envelope during the pre-MS contraction, when magnetic fields several orders of
magnitude stronger were present.  The minimal poloidal magnetic field that pushes the core toward solid rotation
is well below the threshold of detectability if transported to the surface of the star.  We note that our
assumption of solid rotation in radiative material depends on the relative weakness of
additional processes that might source differential rotation, such as the damping of gravity waves 
that are excited in adjacent convective zones.

We are led to a simpler model for the angular momentum profile in a post-MS star:  nearly solid
rotation in the core, as enforced by large-scale Maxwell stresses, combined with inhomogeneous rotation 
in the convective envelope.  The details of a dynamo operating around the core-envelope boundary,
the ensuing flow of magnetic helicity into the core, and the conditions under which the dynamo
shuts off and core and envelope decouple, are addressed in Paper II.  Here our focus is on the 
angular velocity profile of the envelope, especially as it is influenced by the Coriolis force.

Convective Reynolds stresses appear to transport angular momentum inward in accretion flows \citep{BalbHS1996}.
This effect is demonstrated in stars with extended convective envelopes by \cite{brunp2009} using 
three-dimensional anelastic simulations.  In slowly rotating cases, most relevant to the present 
discussion, the specific angular momentum is found to be independent of radius.  

The rotation frequency of the solar convective envelope is, by contrast, roughly independent of radius 
-- a fact that appears to have influenced treatments of rotation in the post-MS phase.  Although
the solar envelope also covers many scale heights, it has a modest aspect ratio (0.7:1 in radius), 
implying that thermal boundary conditions have a stronger influence on the rotation profile than in a 
star with a deep envelope (e.g. \citealt{browning06}).  The Sun is not especially rapidly rotating by the standards of 
late-type MS stars, although the Coriolis force still has a significant effect on its rotation profile.  
The rotation period $P_{\rm rot} = 2\pi/\Omega$ is about 3 times the local convective time 
$\tau_{\rm con} = \ell_P/v_{\rm con}$ (where $v_{\rm con}$ is the speed of a convective eddy and $\ell_P$ 
the pressure scale height), corresponding to a Coriolis parameter ${\rm Co} \equiv \Omega \tau_{\rm con} \sim 2$.  
Dynamo action may also influence the rotation profile \citep{brun04}.

The most important decision to be made here is how to handle the backreaction of the Coriolis force on the
inward pumping of angular momentum.  We investigate two simple prescriptions:  i) a balance between the Coriolis force and non-radial 
pressure gradient forces that result from the entropy difference between convective downflows and upflows; and ii) an upper limit 
${\rm Co} \sim 1$ in the inner envelope, representing an approximate balance between the Coriolis and inertial forces. 
Both of these prescriptions are compared against asteroseismic fits to {\it Kepler} data for subgiants and 
post-core He flash stars. We find that prescription i) provides a very promising quantitative fit.   

Measurements of surface rotation in subgiants provide independent constraints on the internal angular momentum distribution in post-MS stars.
\cite{schrijver93} have used stellar models to argue for an inconsistency between these surface measurements and the evolution of 
$\sim 1.5\,M_\odot$ subgiants that maintain solid rotation. Adding a more rapidly rotating core, while maintaining solid rotation in the 
envelope, was found to have a negligble impact on the result. The discrepency between models and data was therefore interpreted in terms of 
an additional magnetic wind torque acting on a star as it develops a convective envelope upon leaving the MS.

Here we show that the same inward pumping of angular momentum in the envelope that reproduces the 
{\it Kepler} observations also
causes a factor $\sim 1/2$ reduction in surface rotation speed over the same interval, thereby removing most (but not
necessarily all) of the evidence for a magnetic wind torque.  The calculations presented here, and in Paper II,
neglect any rotational effect of surface magnetic fields during the post-MS phase.

We also extend our considerations to the tip of the RGB and AGB.  Here the Coriolis force has a reduced effect, due to the 
greater expansion of the star ($\gtrsim 10$ times larger than the {\it Kepler} sample analyzed by \citealt{becketal2012}
and \citealt{mossetal2012}).  This leads us to consider more inwardly peaked rotation profiles than the one
suggested by fitting {\it Kepler} data.

A final consideration is binary interaction.   This is especially important for stars which lose
most of their natal angular momentum to a magnetized wind on the MS, corresponding to an initial mass $\lesssim 1.3\,M_\odot$.  
After such a star ascends to the upper RGB or AGB, the inward pumping of angular momentum cannot, by itself, sustain 
${\rm Co}\gtrsim 1$ in the convective envelope.  

A robust hydromagnetic dynamo is maintained during the greatest expansion of the star only if it encounters an
external source of angular momentum.  We consider the interaction of a convecting giant with a companion planet, 
calculating the combined orbital evolution of the planet and the change in the angular momentum of the star.  
An interesting detail of this evolution is the excitation of orbital eccentricity by the gravitational 
quadrupole associated with large-scale convective eddies in the giant.  In the case of a planetary companion, 
this effect is much more pronounced than when the companion is a neutron star \citep{phinney92}, and we show 
that it extends the range of orbital periods within which the planet gives up its orbital angular momentum to the star.

\subsection{Plan of the Paper}

We investigate the rotation profile of a deep convective envelope in Section \ref{s:pump}, with a focus on
the backreaction of the Coriolis force on angular momentum pumping.  Various angular velocity profiles are compared
against the {\it Kepler} measurements of core rotation for subgiants and post-He flash stars in Section \ref{s:1.5Msun}.
The effect of angular momentum pumping on the evolution of surface rotation of subgiants is examined in Section
\ref{s:spindown}.

In Section \ref{s:rotation}, we consider the rotational evolution of model stars of initial
mass 1 and $5\,M_\odot$, including the effects of convective pumping, angular momentum loss to a wind, and 
planetary interaction.   
Details of the evolution of the orbit of a planet around an expanding giant are given in Section \ref{s:orbital_evolution}, including 
drag from the tide raised on the star and the excitation of eccentricity by a convective quadrupole.  

Some outstanding questions are addressed in the concluding Section \ref{s:conclusions}.
The appendix gives further details of our calculation of eccentricity growth in the orbit of a companion planet.

\section{Pumping of Angular Momentum in Deep Convective Envelopes:  Backreaction from the Coriolis Force}\label{s:pump}

The deep convective envelope of a giant star divides into an outer part that is slowly rotating (${\rm Co} \lesssim 1$);
and an inner part where ${\rm Co} \gtrsim 1$ and the rotation profile is influenced by the Coriolis force.  

The entire envelope may rotate slowly near maximum expansion, especially if the spin angular momentum of the star
has been strongly depleted by a magnetized wind during the MS phase.  
On the other hand, a star is still compact enough during the early stages of post-MS expansion that $\Omega \tau_{\rm con} \gg 1$
throughout much of the convective envelope.  

We imagine that entropy is approximately conserved over downflows and upflows that are extended
compared with a local scale height.   Some indirect evidence for such a configuration has long been obtained
from one-dimensional models of giant convection, which require a mixing length exceeding a single
scale height \citep{sb91}.

Such a suppression of radial mixing is suggestive of an inwardly peaked rotation profile in the outer, slowly rotating
envelope, corresponding to the conservation of specific angular momentum $j$ by individual convective flows --  
as indeed is obtained numerically by \cite{brunp2009}.  Rotation
profiles intermediate between $\Omega(r) \propto r^{-2}$ and solid rotation could clearly be obtained from limited mixing,
but for definiteness we focus here on the simple case $\partial j/\partial r \rightarrow 0$ where 
${\rm Co} < 1$.  Obtaining the angular velocity profile in the inner envelope depends on a more detailed analysis
of the Coriolis force, which we give below.

We consider the evolving partition between fast and slow rotation
in Section \ref{s:partition}, and how it depends on the angular momentum retained (or gained) by the star in Section
\ref{s:rotation}.
Given a flat distribution of specific angular momentum in the outer envelope, an intermediate-mass star generally sustains an
inner zone where ${\rm Co} \gtrsim 1$, even near the tip of the RGB or AGB;
but a star which spins down on the MS must gain angular momentum from a planetary or
stellar companion.

\subsection{Extended Upflows and Downflows with Quasi-Geostrophic Balance}\label{s:extend}

When the inner zone with ${\rm Co} \gtrsim 1$ covers a wide range of radius, we can look for a power-law scaling of 
the rotation frequency $\Omega$ with spherical radius.    The stellar envelope is nearly adiabatic and spherically stratified.
In this inner zone of fast rotation ($\Omega r\sin\theta \gg v_{\rm con}$) about an axis $\hat z$ ($\theta = 0$),
the steady vorticity equation reads (e.g. \citealt{balbus09})
\be\label{eq:vort}
r\sin\theta {\partial\Omega^2\over\partial z} \simeq -{1\over C_p \rho r}{\partial P\over \partial r}
{\partial S\over\partial \theta} =  {g(r)\over C_p r} {\partial S\over \partial \theta}.
\ee
Here $P$ and $\rho$ are pressure and density, $g(r) = GM(r)/r^2$ is gravity,
and the specific entropy is $S = [\mu(\gamma~-~1)]^{-1}\ln(P/\rho^\gamma)$,
approximated here as that of an ideal gas with mean atomic weight $\mu$ and specific heat $C_p$ ($C_v$)
at constant pressure (volume).

The `thermal wind' approximation used in Equation (\ref{eq:vort})
holds because $|\partial P/\partial \theta| \ll |\partial P/\partial \ln r|$.  
In the upper, slowly rotating part of the envelope, the latitudinal pressure gradient
is sourced by the entropy difference between upflows and downflows,
and is smaller by a factor $(v_{\rm con}/c_s)^2$ than the radial gradient.  
Our focus here is on a rapidly rotating solution to Equation (\ref{eq:vort}), where the angular pressure
gradient which sources $\partial S/\partial\theta$ is proportional to $\Omega^2$.

When the gravitating mass in the inner envelope is dominated by the stellar core, gravity $g(r) \propto r^{-2}$.  
From Equation (\ref{eq:vort}) we obtain the ansatz
\be\label{eq:omsq}
\Omega^2(r,\theta) = \Omega^2(R_c)\left({r\over R_c}\right)^{-3}  f(\cos\theta).
\ee
Here $R_c$ is the radius inside of which ${\rm Co} \gtrsim 1$.  

One subtlety here is that the stellar mass profile is not as centrally concentrated during the early
subgiant expansion.  This generally leads to a shallower scaling of $\Omega$ with $r$:
taking $g(r) \propto r^{-\beta}$, with $\beta < 2$, we find $\Omega(r) \propto r^{-(1+\beta)/2}$.  A fit to the gravity profile in 
the inner convective envelope over the range of expansion that is probed by the {\it Kepler} asteroseismic data gives
$\beta \sim 1$, corresponding to $\Omega(r) \propto r^{-1}$.  The monopolar scaling of $g$ is approached
following the first dredge-up, and is maintained near 
the tips of the RGB and AGB.

We imagine that the rotation profile organizes at ${\rm Co} \gg 1$ into
axially symmetric rolls.  In other words, the strong latitudinal dependence of the Coriolis force is
hypothesized to translate into a persistent latitudinal entropy gradient.  If, furthermore, the entropy
is conserved along radially extended convective plumes, then we can decompose the 
entropy profile into low-order spherical harmonics, $S(r,\theta,\phi) = S_0 + \delta S_0 P_\ell(\cos\theta)$.  
Substituting this into Equation (\ref{eq:vort}) along with Equation (\ref{eq:omsq}), and normalizing
$f = 1$ at $\theta = 0$, we get
\ba
  f &=& 1\quad (\ell = 2); \quad\quad f = {21\over 20}\cos^2\theta - {1\over 20}\quad (\ell = 4);\nn
  f &=& {1\over 280}\left(495\cos^4\theta - 234\cos^2\theta + 19\right) \quad (\ell = 6).
\ea

Angular and radial gradients in $\Omega$ are generally comparable in this type of flow (excepting the simplest
case of a quadrupolar entropy pattern).  We note that rotation is fastest at the pole for a given
$\ell$ (corresponding to an `anti-solar' rotation profile), but that the opposing gradient can be
obtained by superposing $\ell = 2$ with any higher harmonic.  A small anti-rotation is also present in a narrow
latitudinal band when the entropy follows a single spherical harmonic.

A caveat here involves our neglect of the Lorentz force, which may have a considerable influence
near the boundary of a convection zone, but does not obviously compete with the right-hand side of
equation (\ref{eq:vort}) within a deep convective layer.  In particular, the magnetorotational instability
(MRI) is ineffective in the outer part of the envelope where ${\rm Co} < 1$, because the growth time of a MRI mode
is longer than the convective time.

\subsection{Rapid Mixing Between Upflows and Downflows}

We now turn to consider the consequences of rapid mixing between upflows and downflows
for the rotation profile in parts of the envelope that reach ${\rm Co} \sim 1$.  This provides
a useful comparison with the results of Section \ref{s:extend}, and demonstrates how the
radial angular velocity gradient is tied to the presence of radially extended convective flows.  
We show that the downward advection of angular momentum cannot push the flow to ${\rm Co} \gg 1$ 
in the inner part of an extended envelope, where $\tau_{\rm con}$ {\it decreases} inward.  This 
result has a simple interpretation: the competition between the Coriolis force and the inertial force
limits ${\rm Co}$ to a value of the order of unity.

In the mixing length approximation, the entropy difference between upflows and downflows
is related to the speed of a convective eddy by (e.g. \citealt{bethe90})
\be
\delta S = {\delta h\over k_{\rm B}T} = {C_p\over C_p-C_v}{v_{\rm con}^2 \over k_{\rm B} T}.
\ee
Here $\delta h$ is the perturbation to specific enthalpy and $T$ is temperature.  Then the
angular entropy gradient can be estimated as $\partial S/\partial\theta \sim (r/\ell_P)\delta S$.
Because it turns out that ${\rm Co} = O(1)$, we cannot take Equation (\ref{eq:vort}) literally, but
can still use it to estimate the magnitude of the shift of angular velocity across a scale height:
\be\label{eq:mix}
\delta(\Omega^2) \sim {g(r)\over \ell_P} {v_{\rm con}^2\mu\over k_{\rm B} T} = 
\left({v_{\rm con}\over \ell_P}\right)^2.
\ee

We look for a rotation profile that diverges inward, so that the constant term in $\Omega$ can be neglected.  
Then we find
\be
\Omega(r) \sim {v_{\rm con}(r)\over \ell_P}.
\ee
As advertised, the Coriolis parameter does not rise much above unity.  (Faster rotation can, of course,
still be obtained by superposing uniform rotation on the flow.)

For example, in a nearly adiabatic envelope with $g(r) \propto r^{-2}$, 
the speed of an eddy scales as $v_{\rm con} \propto (\rho r^2)^{-1/3}$.
Then $\rho(r) \propto r^{-3/2}$ implies $\Omega(r) \propto r^{-7/6}$ for $\gamma = 5/3$,
which is shallower than the profile (\ref{eq:omsq}).

\subsection{Model $\bar\Omega(r)$ Profiles}

A simple parameterization of the rotation profile which incorporates these lessons may be constructed
as follows.  Where the Coriolis force can be neglected, at $r > R_c$, we consider a flat distribution 
of specific angular momentum,
\be\label{eq:omouter}
\bar\Omega(r) = \bar\Omega(R_c) \left({r\over R_c}\right)^{-2};  \quad r > R_c.
\ee
Here $\bar\Omega$ is the angular velocity averaged over a spherical shell.  If the angular momentum of 
the star is small enough that $\bar\Omega(r) \tau_{\rm con} < 1$ throughout the envelope, then we identify 
$R_c$ with the base of the envelope,
\be
R_c = R_{\rm benv};  \quad (\Omega \tau_{\rm con} < 1\;{\rm at} \; R_{\rm benv}).
\ee

When the star has more angular momentum, we maintain the profile (\ref{eq:omouter}) in the
outer part of the convection zone, and consider
\ba\label{eq:ominner}
({\rm I}) \quad\quad \bar\Omega(r) &=& \bar\Omega(R_c) \left({r\over R_c}\right)^{-\alpha} \nn
&=& {{\rm Co}_{\rm trans}\over \tau_{\rm con}(R_c)} \left({r\over R_c}\right)^{-\alpha};  \quad
R_{\rm benv} < r < R_c\nn
\ea
in the inner envelope.  Here ${\rm Co}_{\rm trans} = O(1)$. As the angular momentum increases $R_c$ moves outward, and may reach the outer 
radius of the star, in which case ${\rm Co}$ is larger than unity everywhere in the envelope.

Following the above discussion, we will explore power-law indices $1 < \alpha < 3/2$,
with the preferred value depending on the radial scaling of gravity.
Such a rotation profile corresponds  to transport of energy by radially extended, adiabatic plumes.

\begin{figure}[!] 
\epsscale{1.0}
\plotone{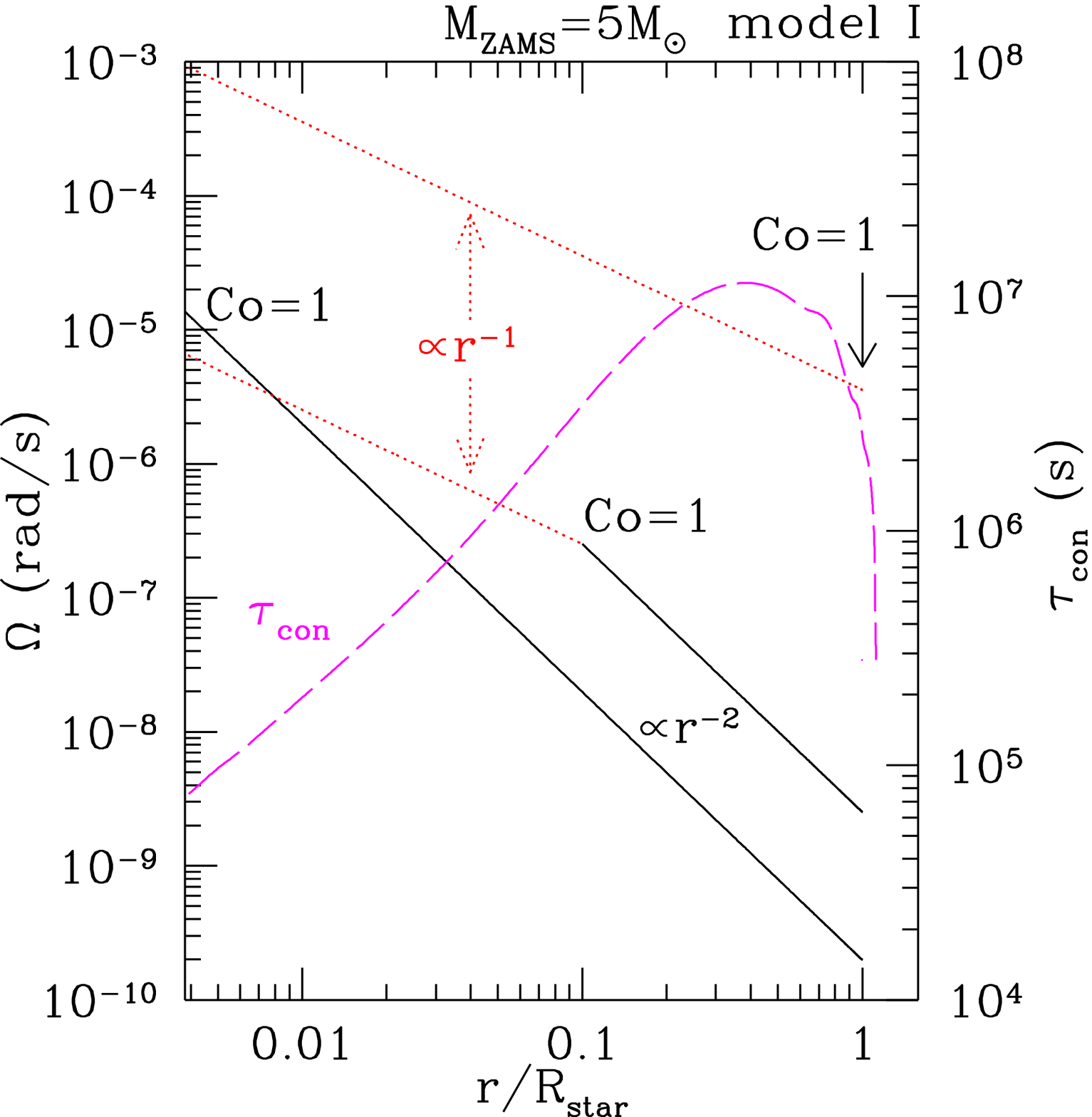}
\vskip .2in
\plotone{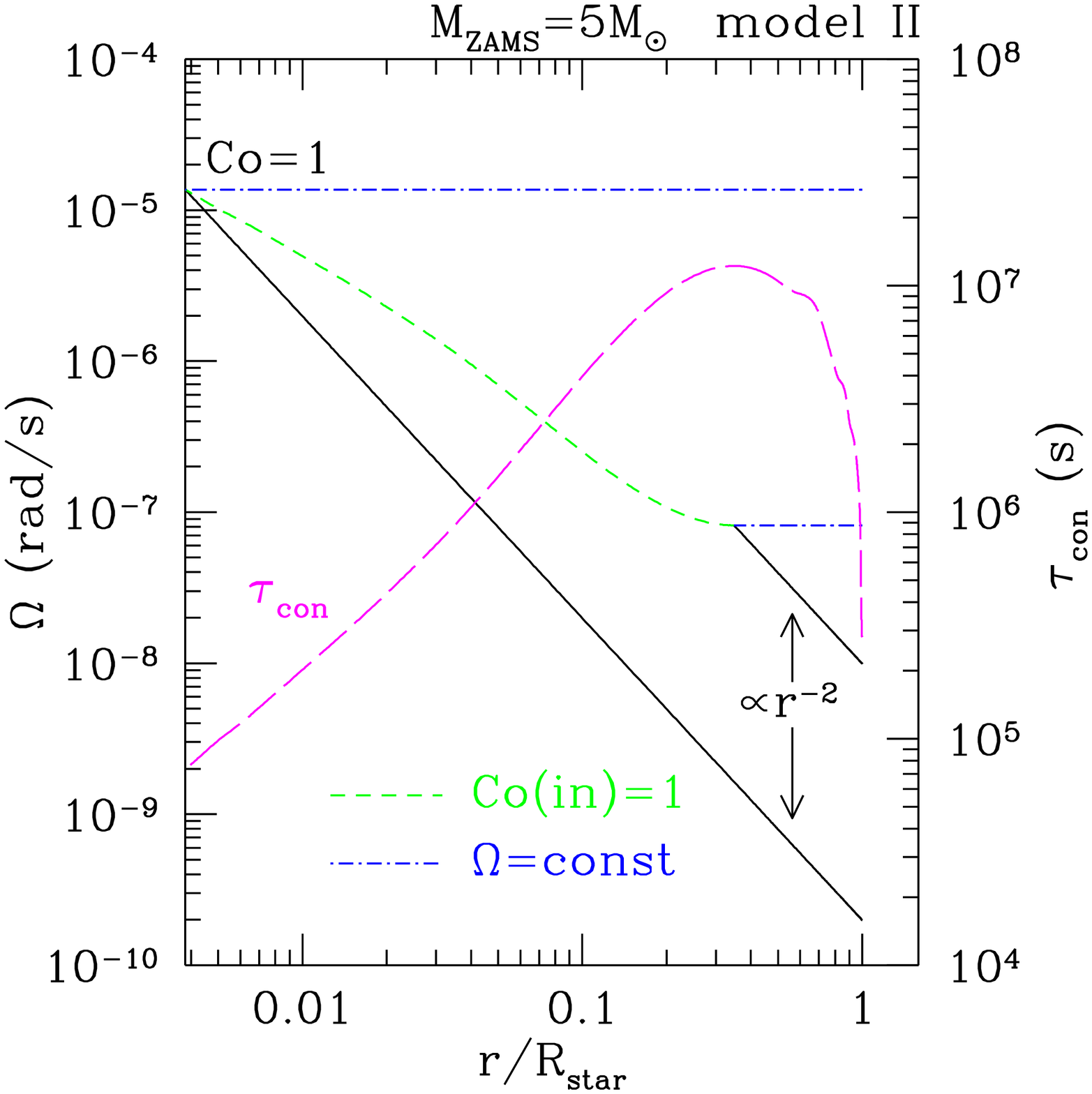}
\caption{Examples of rotation models I (top panel) and II (bottom panel), both taken from the $5\,M_\odot$ model
near the tip of the AGB.  Each panel shows several rotation profiles corresponding to
different total stellar angular momenta.  The magenta curve shows, for reference, the profile of $\tau_{\rm con}(r)$.
Black solid lines represent segments where $\Omega(r)\propto r^{-2}$ and red dotted lines $\Omega(r)\propto r^{-1}$. 
Top panel:  as $J_\star$ is increased, the transition radius $R_c$, for which ${\rm Co}(R_c)=1$, moves outward.
Bottom panel: profiles from bottom to top represent models with increasing $J_\star$, but now when $R_c$
reaches the radius of maximum $\tau_{\rm con}$, we begin to add a uniform $\Omega$ component (the last term in
Equation (\ref{eq:ominnerb})).  This prevents the formation of an outer zone of positive $\partial\Omega/\partial r$.
Black curve: constant $j$, reaching ${\rm Co}=1$ at the base of the envelope; green curve: $\Omega(r) = v_{\rm con}/\ell_P$;
blue curves:  constant $\Omega$ component that grows with increasing $J_\star$.  
{\it Kepler} asteroseismic data favor model I over model II.}
\vskip .2in
\label{fig:rotational_profiles_limiting_cases}
\end{figure}

A second profile, based on mixing length theory, is obtained by setting ${\rm Co}$ to a constant everywhere inside the radius $R_c$,
\be\label{eq:ominnerb}
({\rm II}) \quad\quad \bar\Omega(r) = 
{{\rm Co}_{\rm trans}\over \tau_{\rm con}(r)} + \Omega_0; \quad R_{\rm benv} <  r < R_c.
\ee
The most important difference with profile I is that ${\rm Co}$ remains pegged at ${\rm Co}_{\rm trans} \sim 1$
in the absence of the constant term.   Setting $\Omega_0 = 0$
strictly limits the angular momentum which the envelope may contain, a bound which is easily violated
during core helium burning.   We also require 
that $R_c$ is bounded above by the radius at which $\tau_{\rm con}$ reaches a maximum;  this prevents the formation
of a layer below the photosphere in which $\bar\Omega$ increases outward.   See Figure
\ref{fig:rotational_profiles_limiting_cases}.

\subsection{Latitudinal Differential Rotation}\label{s:latdiff}

The latitudinal shift in rotation frequency (e.g. between pole and equator) has an important influence on the
growth of magnetic helicity in the growing core of an RGB or AGB star (Paper II).  In rotation model I,
where ${\rm Co} > 1$ is maintained by radially extended convective flows, we deduce a strong latitudinal shift:
$\Delta\Omega/\bar\Omega \equiv (\Omega_{\rm eq}-\Omega_{\rm pole})/\bar\Omega \sim 1$.
A strong shift is also maintained in the mixing length approximation (rotation model II), as long as the envelope
rotates rapidly enough to maintain ${\rm Co} \sim 1$ at the base of the envelope.  It is
seen in the more rapidly rotating simulations of \cite{brunp2009}.

We note that weaker latitudinal differential rotation 
is expected in a compact convective envelope with $v_{\rm con} \ll \Omega r\sin\theta$, as
is encountered in the Sun, $\Delta\Omega/\bar\Omega \propto {\rm Co}^{-2}$.

\subsection{Matching of Rotation between \\ Core and Convective Envelope}\label{s:couple}

The rotation of core and envelope are easily coupled to each other when the inner envelope retains enough angular 
momentum to sustain a large-scale hydromagnetic dynamo.   The details of this coupling depend on whether the core
grows, or recedes, as measured in the radial mass coordinate.  When the core is growing, the instantaneous state
of the dynamo is the main consideration, because magnetic fields that are amplified near the core-envelope boundary
are advected downward.  When the core is receding in mass (as it does when a star first ascends the giant branch during the first dredge-up phase), 
the coupling also depends on a prior dynamo process that implanted a magnetic field in the core material.  

\begin{figure}[!] 
\epsscale{1.0}
\plotone{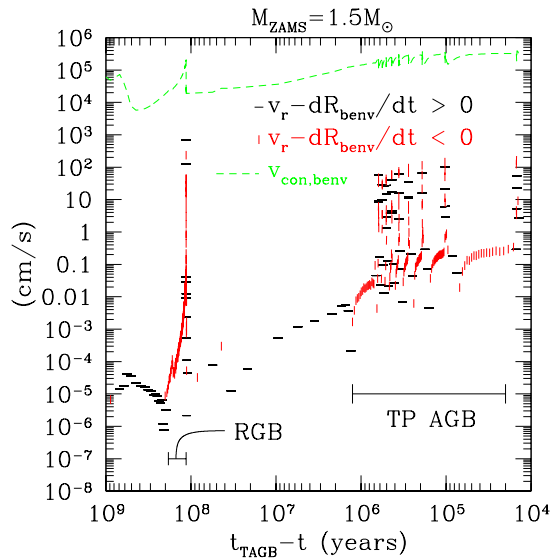}
\vskip .2in
\plotone{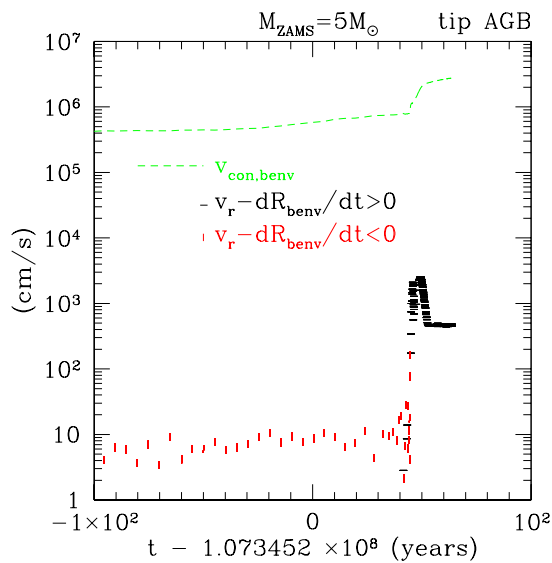}
\caption{Flow speed of stellar material with respect to the core-envelope boundary, compared with convective speed in the inner envelope. 
Upper panel:  $1.5\,M_\odot$ model from the early subgiant expansion to the thermally pulsating AGB. 
Bottom panel:  $5\,M_\odot$ model at the tip of the AGB.
The logarithmic abscissa in the upper panel is referenced to the time $t_{\rm TAGB}$ of envelope ejection.
The time coordinate in the lower panel has been shifted to focus on the final few hundred years.}
\vskip .2in
\label{fig:drift}
\end{figure}

A key consideration is the magnitude of the
drift speed $v_r - dR_{\rm benv}/dt$ of stellar material with respect to the core-envelope
boundary (averaged if necessary over thermal pulsations, which are present on the terminal AGB).  Solid rotation is enforced
in the outer core during downward drift if the radial Alfv\'en speed is larger than this drift speed.   This requires
that the large-scale poloidal magnetic field that is generated in the inner envelope be strong enough that
\be\label{eq:couple}
{B_r^2\over 4\pi \rho(R_{\rm benv}) v_{\rm con}^2} \gtrsim {(v_r-dR_{\rm benv}/dt)^2\over v_{\rm con}^2}.
\ee
We show in Figure \ref{fig:drift} both the drift speed relative to the core-envelope boundary, and
the convective speed in the inner envelope (as evaluated using mixing-length theory),
during the post-MS evolution of a $1.5\,M_\odot$ model star.  The same quantities are also shown 
for a $5\,M_\odot$ star near the tip of the AGB.  (Further details of these models are described in Sections \ref{s:1.5Msun} and \ref{s:rotation}.)

One sees that during the early subgiant expansion of the $1.5\,M_\odot$ model, the remnant poloidal magnetic field
in the outer core need only retain an energy density that exceeds $\sim 10^{-16}-10^{-15}$ of the instantaneous 
convective energy density in order to compensate changes in rotation induced by the evolving mass profile.
Magnetic fields $\sim 10^2$-$10^3$ times stronger are needed to compensate
a latitude-dependent convective stress (Paper II).

The envelope decreases more rapidly in mass near the tip of the RGB and AGB,
and the magnetic energy density that is self-consistently generated during these phases must rise to about
$\sim 10^{-8}-10^{-6}$ of the convective energy density.  
We conclude that magnetic decoupling of core from
envelope is most easily achieved during the brief phase where the envelope is expelled.   



In Section \ref{s:1.5Msun}
we concentrate on the comparison with {\it Kepler} asteroseismic data, which only covers post-MS stars more
compact than $\sim 10 R_\odot$.  These subgiants and core He burning stars evolve on a relatively long timescale ($\sim 10^2$ times 
the duration of the thermally pulsating AGB phase).  The results shown in
Figure \ref{fig:drift}
strongly suggest that the rotation of core and envelope remain well coupled
in these stars.  

We therefore take the core to rotate as a solid with the mean rotation at the base of the envelope,
\be\label{eq:omcore}
\Omega_{\rm core} = \Omega_{\rm benv} \equiv \bar\Omega(R_{\rm benv}).
\ee
An explanation for the rotation behavior of the {\it Kepler} sample must be found
in the redistribution of angular momentum within an extended convective envelope.

\begin{figure}[!] 
\epsscale{1.0}
\plotone{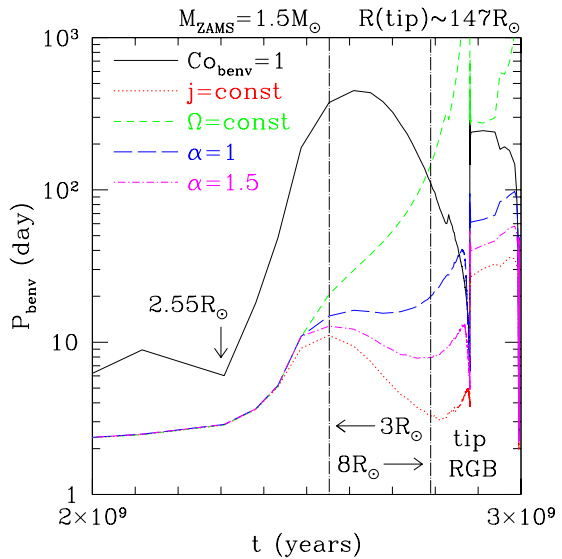}
\caption{Rotation period at the base of the convective envelope, in a $M_{\rm ZAMS}=1.5M_\odot$ model star, as a function of time during the RGB
and early core He burning phases.  Curves correspond to different rotation profiles.  Black line:  rotation period
$P_{\rm benv}$ corresponding to ${\rm Co_{benv}} = 1$ (with $\tau_{\rm con}$ evaluated
in the mixing-length approximation).  Green dashed line:  solid rotation.  Red dotted line:  uniform
specific angular momentum (even where ${\rm Co} > 1$).  Blue long-dashed line:  rotation model I given
by Equations (\ref{eq:omouter}), (\ref{eq:ominner}) with $\alpha = 1$. Magenta dashed-dotted line: rotation model I with $\alpha = 3/2$.  Vertical black lines show the age range probed by the {\it Kepler} subgiant measurements,
with corresponding stellar radius.}
\vskip .2in
\label{fig:rotation_profiles}
\end{figure}

\section{Comparison of $1.5\,M_\odot$ Model Star \\ with {\it Kepler} Data}\label{s:1.5Msun}

We begin by calibrating the rotation models described by Equations (\ref{eq:omouter}) and 
(\ref{eq:ominner})-(\ref{eq:omcore}) against the core rotation periods of $1-2\,M_\odot$ subgiants and post-core He
flash stars as measured by \cite{becketal2012} and \cite{mossetal2012}.  For ease of comparison, we 
use the same stellar model as \cite{CantMBCP2014}, namely a $1.5\,M_\odot$ star of solar
metallicity with zero-age main sequence (ZAMS) equatorial rotation speed 50 km s$^{-1}$.  

Our calculation also employs the 
1D stellar evolution code MESA (\citealt{paxtetal2011}, version 5527),
but with the built-in prescriptions for angular momentum transport turned off.   Instead, the 
rotation profile is `patched' onto the sequence of stellar models.  

The total spin angular momentum of the
star is evolved self-consistently in response to mass loss, using the prescription described
in Section \ref{s:angloss}.  Magnetic wind torques are neglected, for the reasons described
in Section \ref{s:spindown}.  The net effect of mass loss is found to be modest up to the onset of core
He burning:  a factor $\sim 2$ reduction in angular momentum during the RGB phase.  

\begin{figure}[!] 
\epsscale{1.0}
\plotone{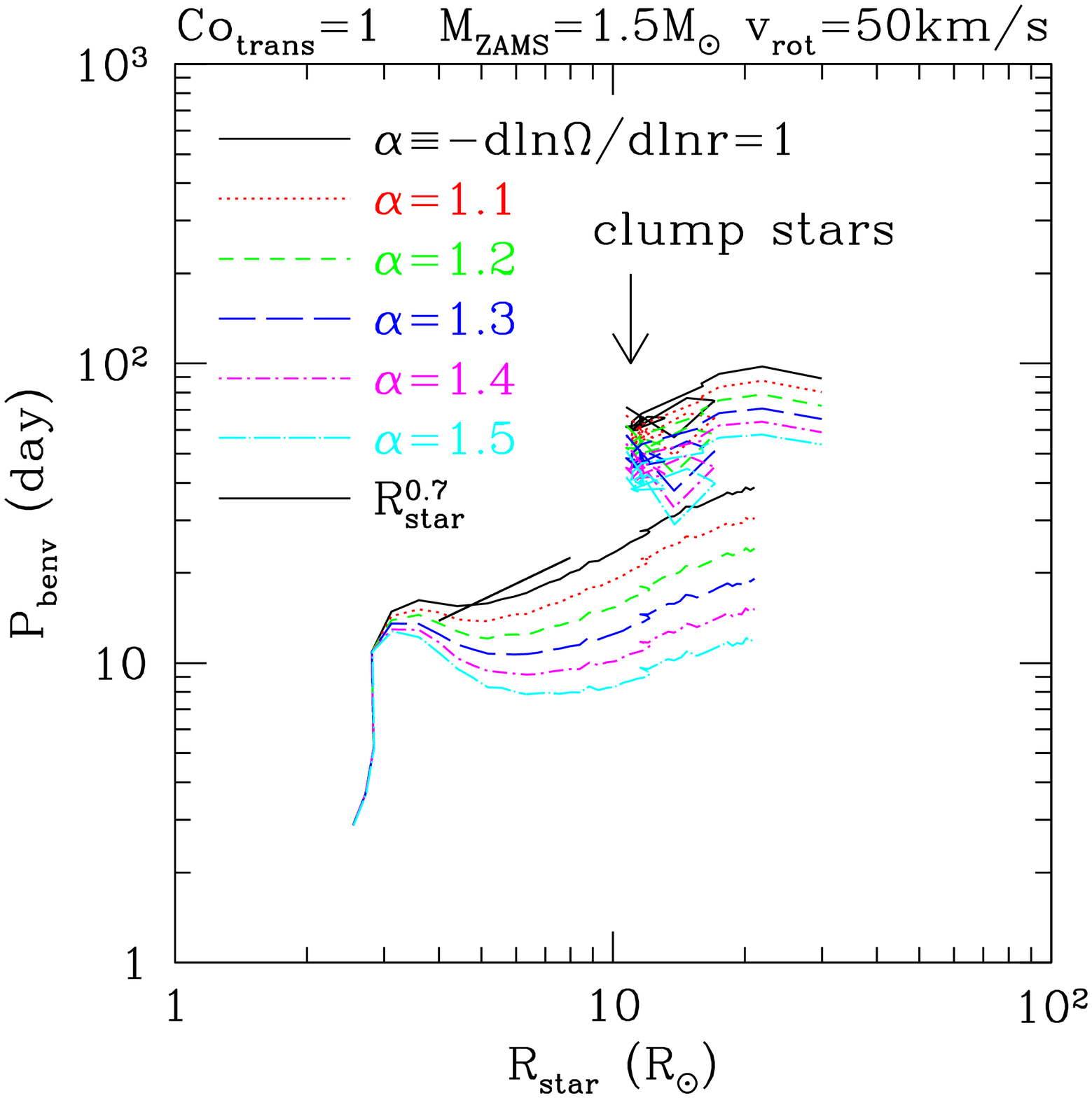}
\vskip .2in
\plotone{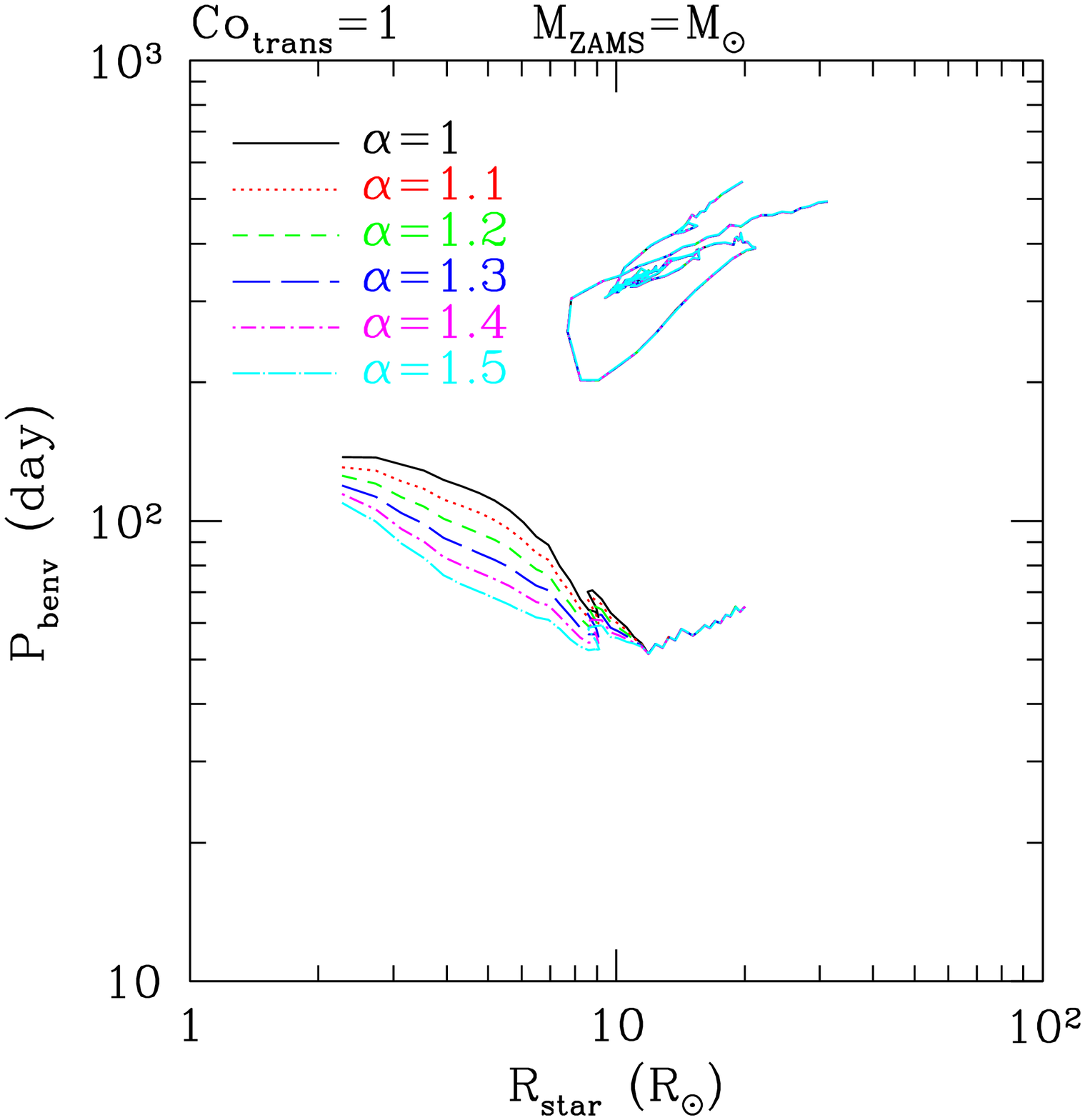}
\caption{Rotation period at the base of the convective envelope during the early giant expansion
and the core He burning phase of a $1.5\,M_\odot$ star (upper panel) and a 1 $M_\odot$ star (lower panel). 
Rotation model I given by Equations (\ref{eq:omouter}), (\ref{eq:ominner}) with ${\rm Co}_{\rm trans} = 1$ and 
varying inner index $\alpha$ (here $\Omega(r) \propto r^{-\alpha}$ inside the radius $R_c$ where  
${\rm Co} = {\rm Co}_{\rm trans}$). 
The 1.5$ M_\odot$ star is given a ZAMS equatorial rotation speed 50 km s$^{-1}$, while the 1 $M_\odot$ 
star is given the solar rotational angular momentum.
Diagonal black line:  fit to {\it Kepler} subgiant data from \cite{CantMBCP2014}.
The radius at which majority of helium clump stars are found is marked in the upper panel.}
\vskip .2in
\label{fig:Kepler_comparison}
\end{figure}

\begin{figure}[!] 
\epsscale{1.0}
\plotone{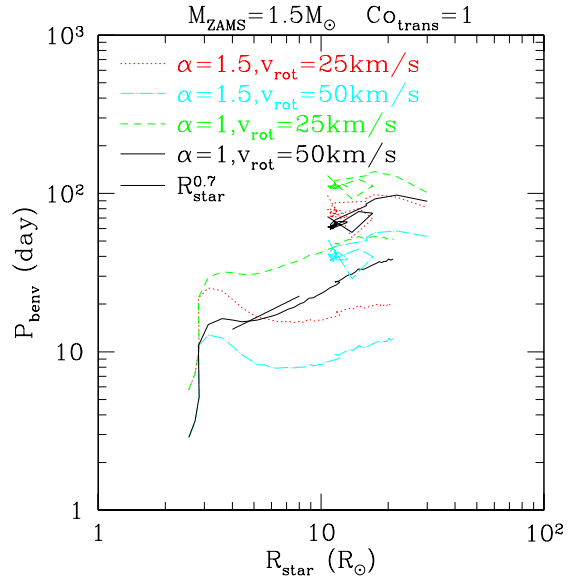}
\caption{Same as upper panel of Figure \ref{fig:Kepler_comparison}, but now for various initial rotation speeds of the 
$1.5\,M_\odot$ star, and two selected scalings for $\Omega(r)$ in the rapidly rotating inner zone.}
\vskip .2in
\label{fig:Kepler_comparison2}
\end{figure}

\begin{figure}[!] 
\epsscale{1.0}
\plotone{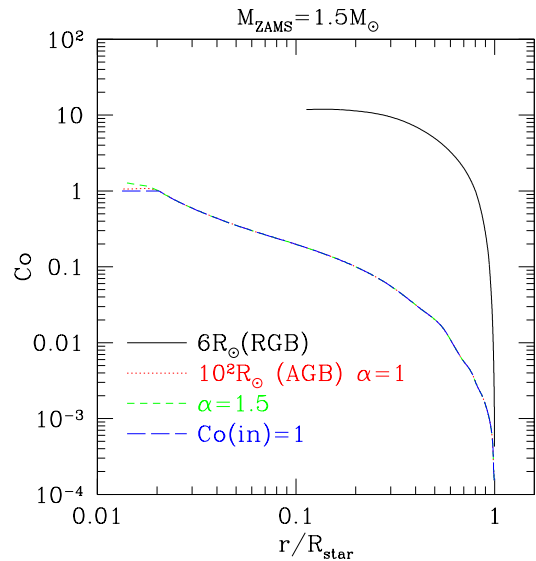}
\caption{Profile of Coriolis parameter with radius during two phases of the evolution of our $1.5\,M_\odot$ model
with initial $v_{\rm rot} = 50$ km s$^{-1}$.  Black curve: subgiant (stellar radius $6\,R_\odot$) with inner rotation 
index $\alpha=1$ in Equation (\ref{eq:ominner}).  Colored curves: near the tip of the AGB (stellar radius 
$100\,R_\odot$) with uniform specific angular momentum in outer envelope, and inner rotation profile corresponding
to either $\alpha = 1$ or 1.5, or to ${\rm Co}$ saturating at unity.}
\vskip .2in
\label{fig:Co_profile}
\end{figure}

We observe from Figure \ref{fig:rotation_profiles} that the inner envelope must rotate rapidly (${\rm Co}_{\rm benv} > 1$),
even in the case of solid rotation, over the range of expansion probed by the {\it Kepler} data.  The rotation period
$P_{\rm benv} = 2\pi/\Omega_{\rm benv}$ at the base of the envelope would actually decrease with increasing expansion if the entire envelope
could sustain a profile $\Omega(r) \propto r^{-2}$.  

This effect is still present in the less peaked rotation profile I, as given by Equations (\ref{eq:omouter}), (\ref{eq:ominner}) with 
$\alpha = 1$ (or 3/2). Now $P_{\rm benv}$ sits close to the observed range of $10-20$ days.  This period is identified here with the core rotation 
period.

Another important feature is the large jump in rotation period between the red giant and core He burning phases (Figure 
\ref{fig:Kepler_comparison}).  The latter, slower, rotation results from a retraction of the convective envelope combined with 
continuing angular momentum pumping and core-envelope synchronization. (It was previously ascribed by \cite{CantMBCP2014} to the 
conservation of angular momentum by an expanding core that has already decoupled rotationally from the inner envelope.)

A closer look at the effect of the inner rotation index $\alpha$, and the initial rotation speed of the star, is provided by Figures 
\ref{fig:Kepler_comparison} and \ref{fig:Kepler_comparison2}. Here one sees that an index $\alpha$ around $1-1.1$ provides a close agreement 
with the {\it Kepler} measurements for stars more massive than $\sim 1.3\,M_\odot$.  There is some degeneracy between the effects of changing $\alpha$ and initial rotation speed. It is interesting to note that this profile agrees well with the analytic scaling derived
for rotation model I in Section \ref{s:extend} in the case (appropriate to the expansion phase probed by the {\it Kepler} data) where $g(r)
\propto r^{-1}$ in the inner envelope.  Recall that this rotation model represents radially extended upflows and downflows in a geometrically
deep, adiabatic envelope.

A different conclusion emerges for stars close to solar mass.  We see in the bottom panel of Figure \ref{fig:Kepler_comparison} that the core of a $1\,M_\odot$ star without a close planetary companion is predicted to 
spin $\sim 5-10$ times more slowly during subgiant expansion than is seen in our $1.5\,M_\odot$ model.  On the other hand, the core rotation during the He burning phase is only $\sim 3$ times slower, and indeed remains within the range measured by {\it Kepler} for clump stars.  Significant spindown by magnetic wind
torques during the subgiant phase \citep{saders13} would have similar effects.

\begin{figure}[!] 
\epsscale{1.0}
\plotone{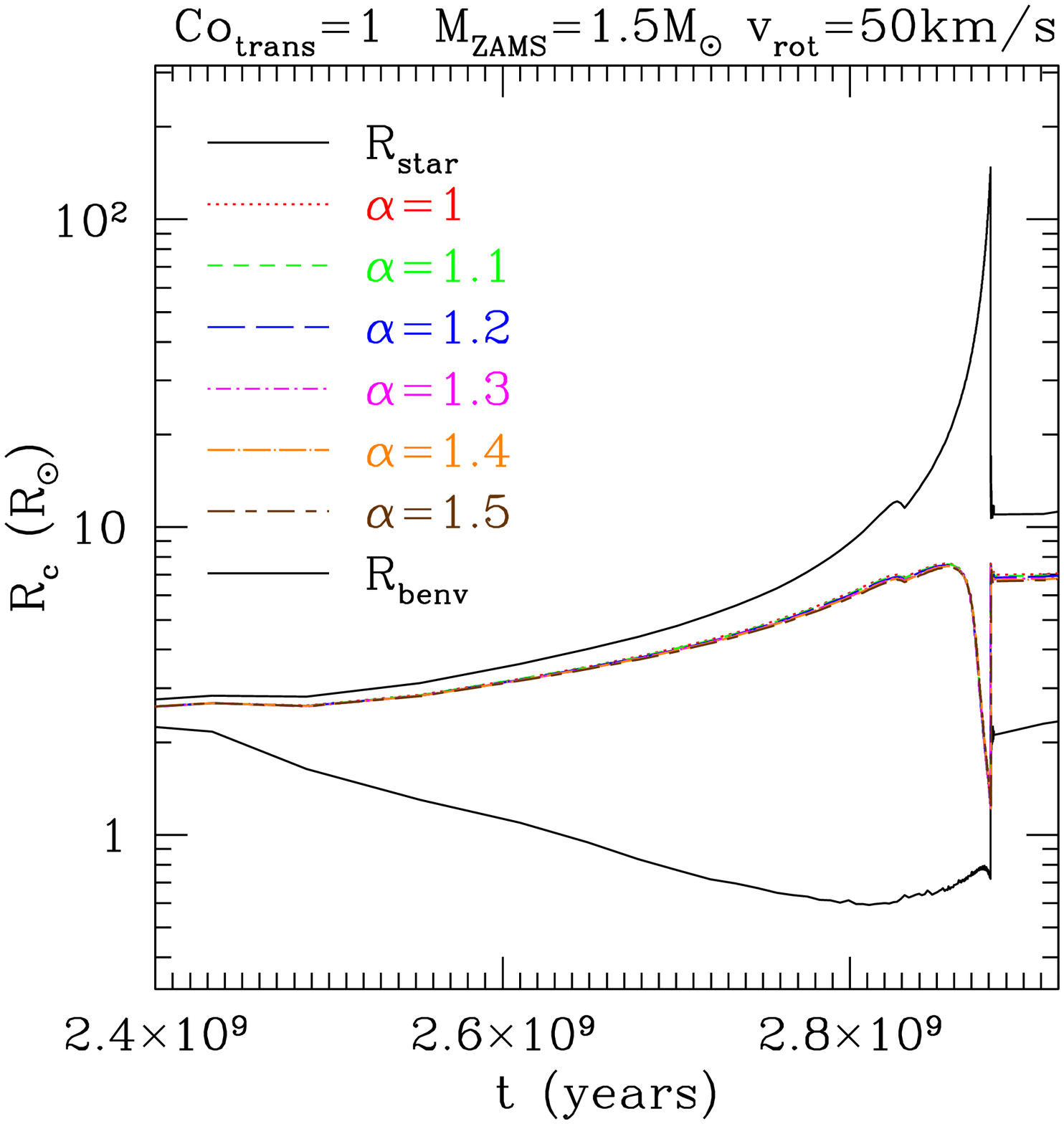}
\vskip .2in
\plotone{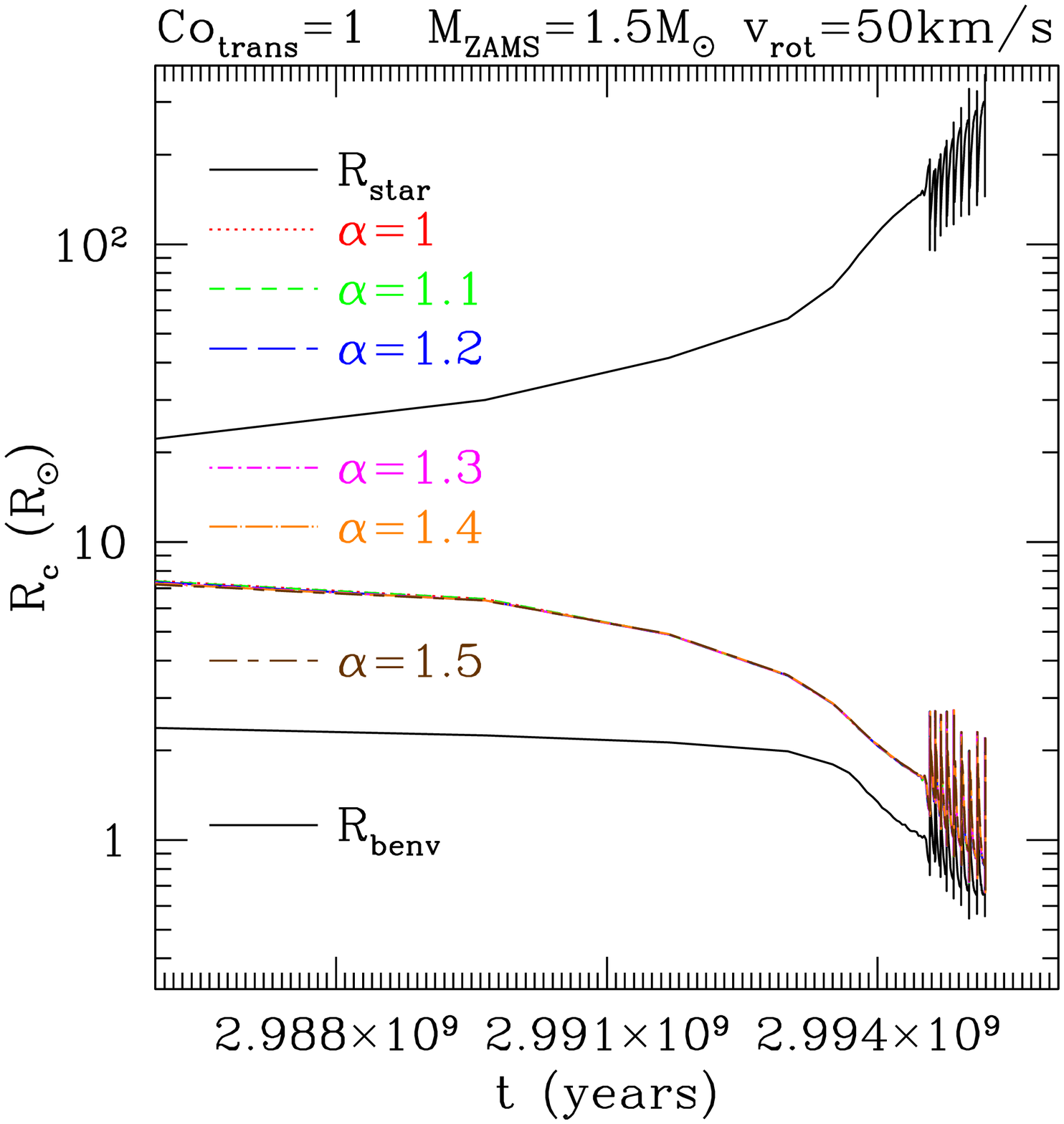}
\caption{Profile of convective envelope during RGB expansion (top panel) and AGB expansion (bottom panel)
in rotation model I.
As the aspect ratio of stellar radius to the base of the envelope grows, the domain within which 
${\rm Co} \gtrsim 1$ shrinks.  The radius $R_c$ marks the transition point between an outer zone
with uniform specific angular momentum, and an inner zone with $\Omega(r) \propto r^{-\alpha}$.  
This transition depends only weakly on the inner index $\alpha$, because most of the angular momentum
is stored in the outer envelope.}
\vskip .2in
\label{fig:Rc_profile}
\end{figure}

\subsection{Transition from ${\rm Co} \gtrsim 1$ to ${\rm Co} \lesssim 1$ throughout \\
Bulk of Convective Envelope}\label{s:partition}

The {\it Kepler} sample of subgiants and core He burning stars probes a different rotational regime than is
experienced by most giants near their maximum expansion.  The {\it Kepler} stars are compact enough that 
we find ${\rm Co} \gtrsim 1$ throughout most of the convective envelope, excepting a relatively thin 
layer below the surface.  Indeed ${\rm Co}$ reaches a considerable value $\sim 10-30$ at the base of 
the envelope (Figure \ref{fig:Co_profile}).  The measured core rotation rates probe the angular velocity 
profile that is sustained in the envelope in a regime of relatively rapid rotation.

Figure \ref{fig:Rc_profile} shows the evolving partition between an outer zone with uniform specific angular
momentum, and an inner zone with angular velocity parameterized by Equation (\ref{eq:ominner}).  
These curves correspond to a star with spin angular momentum close to its birth value, assuming
an equatorial surface rotation speed of $50$ km s$^{-1}$.
Out to a stellar radius $R_\star \sim 10\,R_\odot$, the envelope remains compact enough that ${\rm log}(R_\star/R_c)
\lesssim \log(R_c/R_{\rm benv})$.  Beyond this size, the inner rotation rate is controlled by the rotation
profile at ${\rm Co} \lesssim 1$.  

The spin and magnetization of the WD remnant, as explored in Paper II,
are strongly influenced by the $\Omega(r)$ profile of a {\it slowly} 
rotating convective envelope near the tip of the RGB and AGB.  
Then the gravity profile steepens in the inner envelope, and we
maintain consideration of rotation profiles at ${\rm Co} > 1$ that 
are somewhat steeper than suggested by the fit to {\it Kepler} data:
$\alpha \sim 3/2$ as opposed to $\alpha \sim 1$ in Equation (\ref{eq:ominner}).

\begin{figure}[!] 
\epsscale{1.0}
\plotone{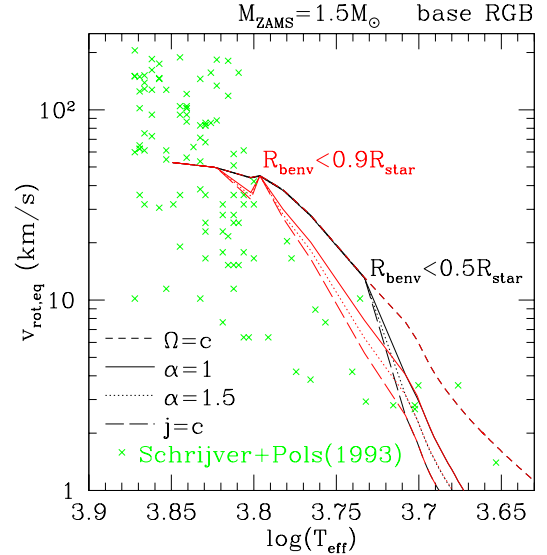}
\caption{Green crosses: observed surface rotation speeds of class IV stars from \cite{schrijver93}, multiplied by
$4/\pi$ to account for unknown inclination, as a function of effective temperature.  We show for comparison
our model of a $1.5\,M_\odot$ star on the early red giant branch, with initial rotation speed $50$ km s$^{-1}$.
Curves correspond to different profiles of rotation in the envelope, combined with solid rotation in the core.  
Solid rotation is enforced in the envelope when the aspect ratio $R_{\rm benv}/R_\star$ between the base 
of the envelope and stellar photosphere is larger than $0.9$ or $0.5$.  The $\alpha=1, 1.5$ curves correspond
to rotation model I.  In the case $\alpha = 1$ (which best reproduces the asteroseismic fits to
{\it Kepler} data) there is a factor $\sim 1/2$ reduction in surface rotation compared with a solid rotator at 
$\log(T_{\rm eff}) < 3.8$.}
\vskip .2in
\label{fig:vsurf}
\end{figure}

\vfill

\subsection{Surface Rotation of Subgiants: \\  Angular Momentum Pumping Versus Magnetized Winds}\label{s:spindown}

The expansion of a star following the MS phase is associated with a drop in surface rotation.  The 
strength of this drop is sensitive to the distribution of angular momentum within the star, as well as
to external torques associated with magnetic activity in the deepening envelope.  \cite{schrijver93} 
and \cite{saders13} have
presented evidence for an inconsistency between solid rotation in the envelope and measurements in the
surface rotation of $1.5-2\,M_\odot$ stars:  the measured rotation is a factor $\sim 2-3$ slower than is
expected for constant stellar angular momentum.  

A similar effect would be imposed by inward pumping of angular momentum. \cite{schrijver93} show
that a fast-spinning core has a negligible effect on the result if the convective envelope maintains solid rotation,
but did not consider envelopes with inward-peaked rotation profiles.  The key point here is that the radiative
core becomes very compact early in the expansion onto the red giant branch, in the sense that its
moment of inertia is a very small fraction of $M_{\rm benv}R_{\rm benv}^2$.  Therefore
inhomogeneous rotation in the envelope has a much larger effect on surface rotation than does a fast-spinning core.

We show in Figure \ref{fig:vsurf} the evolution of the surface rotation in our $1.5\,M_\odot$ model,
for the same set of inhomogeneous rotation profiles that we have previously explored.  
The envelope at high $T_{\rm eff}$ is still relatively thin.  We therefore
impose $\partial\Omega/\partial r = 0$, to mimic the solar rotation profile, 
when the aspect ratio $R_{\rm benv}/R_\star$
rises above a critical value 0.5 or 0.9.  The profile that is favored by the {\it Kepler} core rotation
data, Equations (\ref{eq:omouter}) and (\ref{eq:ominner}) with inner index $\alpha = 1$, shows a factor $\sim 1/2$
reduction in surface rotation compared with the solid rotator.  
This removes a significant part of the
discrepency found by \cite{schrijver93} and \cite{saders13}, but perhaps not all.

\section{Evolving Rotation Profile with Mass Loss and Interaction with a Planet} \label{s:rotation}

We now consider the rotational evolution of a star as it approaches the tip of the RGB and AGB.
Here the probability of interacting with a companion increases significantly.
The Coriolis parameter also drops markedly at the base of the convective envelope.   We maintain our
focus on a two-layered rotation model. 

Two test cases are analyzed in detail:  a star with initial mass $M_{\rm ZAMS} = 1\,M_\odot$, which 
experiences strong spindown on the MS and interacts with a planet; and an intermediate-mass star 
($M_{\rm ZAMS} = 5\,M_\odot$) which remains rapidly rotating at the end of the MS and leaves behind a massive WD.  
This second stellar model is also allowed to interact with a companion.
MESA is used to evolve these stellar models (both of solar metallicity) 
from the ZAMS to the post-AGB phase.

In the case of the solar-mass star, the interaction with a planet was followed from the 
base of the RGB, when the star has expanded to a radius $R_{\star} \sim 10.9R_{\odot}$ (age $t_{10.9} \sim 12.37$ Gyr). 
At that point, the spin angular momentum was set equal to that of the present Sun.
Stellar quantities at times intermediate between the outputed MESA models were obtained by linear interpolation.
Details of the orbital integration and the interaction between planet and star are presented in Section \ref{s:orbital_evolution}.


The equatorial rotation speed of the $5\,M_{\odot}$ star is set to 50 km s$^{-1}$ at the ZAMS,
corresponding to the peak in the measured $v \sin i$ distribution \citep{wolfsdv2007}.  Both the $1\,M_\odot$ and $5\,M_\odot$ 
models are taken to rotate as a solid body on the MS.  Deviations from solid body rotation develop after the formation
of a deep convective envelope, following the approach of Section \ref{s:pump}.

There are instances around the tip of the AGB when the outer envelope splits into several convection zones, 
separated by radiative layers.  The entropy gradient in these radiative zones, although positive, remains small.
We impose a matching of specific angular momentum across them.

\subsection{Loss of Angular Momentum on the RGB and AGB}\label{s:angloss}

Magnetic wind braking is not taken into account in our simulations
during the subgiant or giant phases, for the reasons given in 
Section \ref{s:spindown}.  Then the ejected mass carries the specific 
angular momentum of the stellar photosphere,
\be\label{eq:djdtml}
\dot{J}_{\star} = \frac{2}{3} \bar\Omega(R_{\star}) R_{\star}^2 \dot{M}_{\star}.
\ee
The rate of mass loss on the RGB is handled using the prescription of \cite{reim1975},
\be
\dot{M}_{\star\,\rm Reimers} = 4 \times 10^{-13} \eta_R \frac{\tilde{L}_{\star} \tilde{R}_{\star}}{\tilde{M}_{\star}} \quad M_{\odot}~{\rm yr^{-1}},
\ee
with $\eta_R = 0.5 $.  Here (\textasciitilde) denotes stellar parameters measured in solar units.   On the AGB we use 
\be
\dot{M}_{\star\,\rm Bloecker} = 1.93 \times 10^{-21} \eta_B \frac{\tilde{L}_{\star}^{3.7}\tilde R_\star}{\tilde{M}_{\star}^{3.1}} 
\quad M_{\odot}~{\rm yr}^{-1}
\ee
from \cite{bloe1995}.  Mass loss according to the Bl\"{o}cker formula is turned on when the central H and He
abundances are less than $10^{-2}$ and $10^{-4}$, respectively. 

The choice of AGB mass loss parameter $\eta_B$ is motivated by recent cluster data (e.g. \citealt{Dobbetal2009}).
As we discuss further in Paper II, we set $\eta_B = 0.05$, corresponding to a final WD mass $M_{\rm WD} \sim 0.87M_{\odot}$ (from the
$5M_\odot$ progenitor).
The same wind parameters are used for both the $1\,M_\odot$ and $5\,M_\odot$ models.  The solar-type star loses $\sim0.25M_{\odot}$ on the 
RGB and another $\sim0.21M_{\odot}$ on the AGB for a final WD mass $M_{\rm WD} \sim 0.54M_{\odot}$. 

\begin{figure}[!] 
\epsscale{1.0}
\plotone{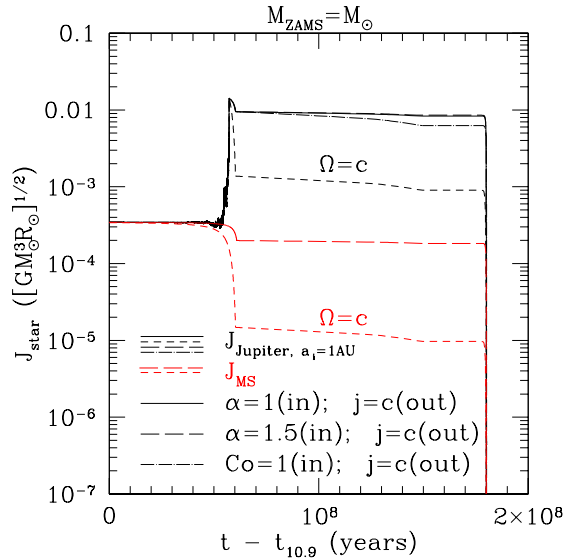}
\caption{Evolution of spin angular momentum of $1\,M_\odot$ model star during RGB, core He burning and final AGB phases. Curves represent 
cases described in the text.  Red lines ($J_{\rm MS}$): no external source of angular momentum.  Black lines ($J_{\rm Jupiter, a_i=1AU}$): 
absorption of a Jupiter-mass planet orbiting initially at $a_i = 1$ AU.  Short dashed lines ($\Omega=c$):  solid 
rotation throughout core and 
envelope.  Other lines correspond to two-layer rotation model, with uniform specific angular momentum in outer envelope,
and varying inner rotation profile.  Solid (long-dashed) curves: $\alpha=1$ ($\alpha=1.5$) in Equation (\ref{eq:ominner});
long-dashed-dotted curve: profile (\ref{eq:ominnerb}) with ${\rm Co} = 1$ in inner zone.}
\vskip .2in
\label{fig:J_star_comparison_RGB_AGB}
\end{figure}

\begin{figure}[!] 
\epsscale{1.0}
\plotone{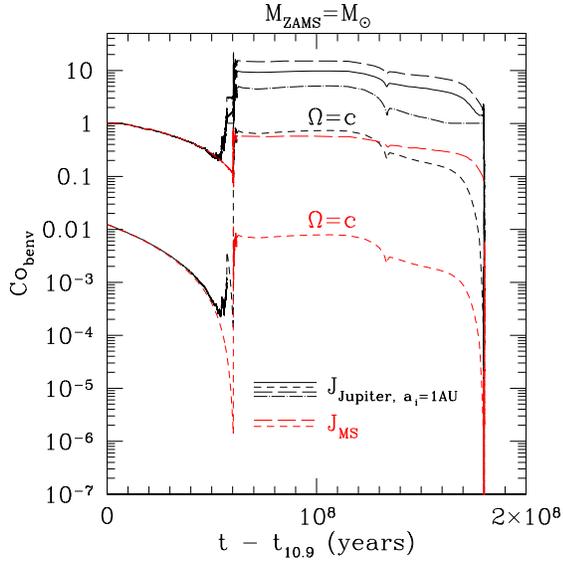}
\vskip .2in
\plotone{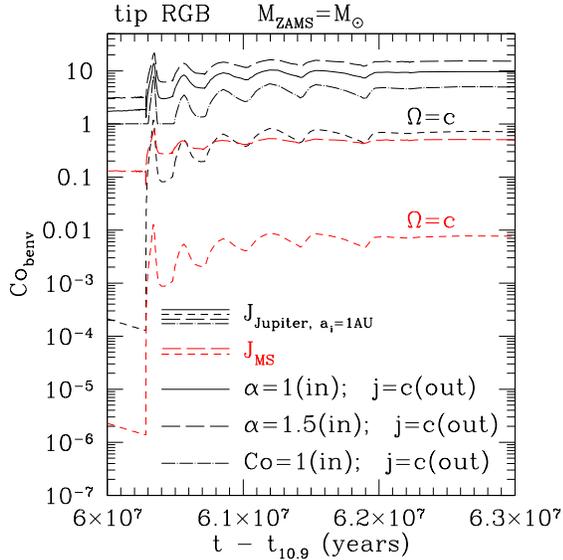}
\caption{Evolution of the Coriolis parameter at the base of the convective envelope. Curves correspond to those in Figure 
\ref{fig:J_star_comparison_RGB_AGB}.
Upper panel shows the RGB, core helium burning phase, and AGB.  Lower panel shows an expanded view of the tip of the RGB and the beginning of
the core He burning phase.}
\vskip .2in
\label{fig:Co_base_comparison}
\end{figure}

\begin{figure}[!] 
\epsscale{1.0}
\plotone{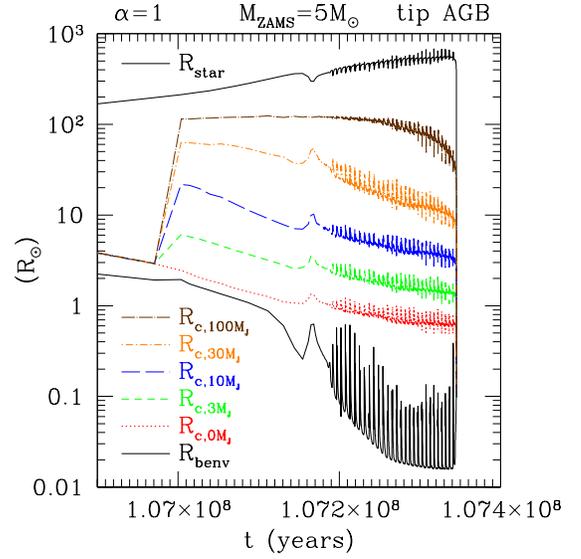}
\vskip .2in
\plotone{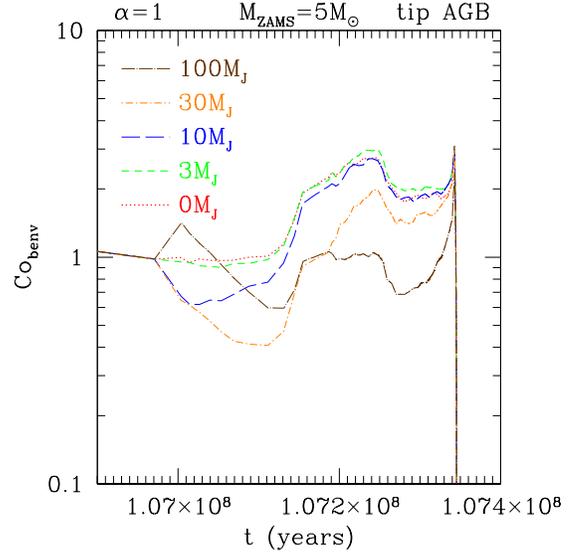}
\caption{Effect on the 5$M_\odot$ model star of the engulfment of a substellar companion of mass
(3,10,30,100)$M_{\rm Jupiter}$ from an initial orbit $a_i = 2$ AU.  Orbital angular momentum of planet is added to
star when it reaches a radius $R_\star = 200R_\odot$ on the AGB, as guided by more detailed tidal evolution calculations.
The physical consequences depend weakly on the epoch of engulfment, if it occurs before the first thermal pulse 
(at $t \sim 1.0717\times10^8$ years), when the envelope dynamo begins to contribute to the magnetic field of the
hydrogen-depleted core.  Upper panel: time evolution of transition radius $R_c$ between outer slowly rotating zone
(${\rm Co} < 1$) and inner rapidly rotating zone (${\rm Co} > 1$).  Bottom panel:  change in $\rm Co_{benv}$ due to the 
added angular momentum.  Rotation profile corresponds to $\alpha=1$ in Equation (\ref{eq:ominner}).
Although $R_c$ increases with increasing angular momentum, $\rm Co_{benv}$ decreases.}
\vskip .2in
\label{fig:R_c_and_Co_benv_for_M_companion}
\end{figure}

\begin{figure}[!] 
\epsscale{1.0}
\plotone{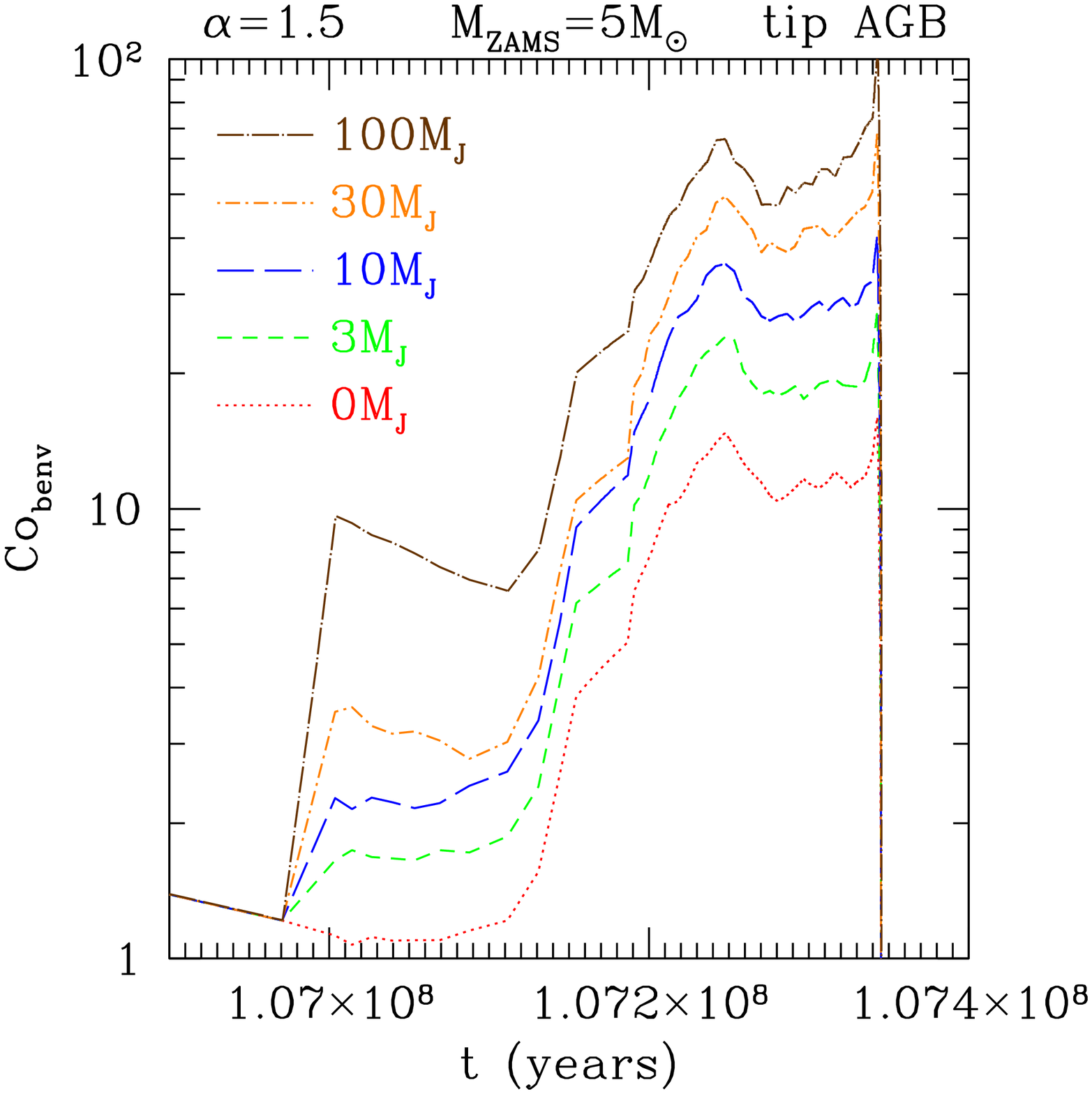}
\caption{Same as bottom panel of Figure \ref{fig:R_c_and_Co_benv_for_M_companion}, but now with steeper inner
rotation profile $\alpha=1.5$ in Equation (\ref{eq:ominner}).  Now ${\rm Co}_{\rm benv}$ increases monotonically
with the added angular momentum.}
\vskip .2in
\label{fig:R_c_and_Co_benv_for_M_companion2}
\end{figure}

\subsection{Rotation Rate of the Inner Envelope and Core} \label{s:RG_rotation}

There is a simple relation between spin angular momentum and core angular velocity $\Omega_{\rm benv}$ 
when the envelope is extended and maintains uniform angular momentum per unit mass,
\be
j(r) \equiv \frac{dJ(r)}{dM(r)} = \frac{2}{3} r^2 \bar\Omega(r) = {\rm const}.
\ee
Given a net stellar angular momentum $J_\star$, and an effective moment of inertia in the convective envelope,
\be
I_{\rm eff} = \frac{2}{3} \int 4\pi r^2 R_{\rm benv}^2 \rho(r) dr = {2\over 3}R_{\rm benv}^2 (M_\star - M_{\rm benv}),
\ee
we obtain 
\be \label{eq:omega_surface}
\Omega_{\rm benv}  = \frac{J_{\star}}{I_{\rm eff} + I_{\rm core}}.
\ee
Excepting near the point where the envelope has been ejected at the end of the AGB, 
the moment of inertia of the core can generally be neglected, $I_{\rm core} \ll I_{\rm eff}$. 

The angular momentum stored in the core is enhanced by a factor $\sim (R_\star/R_{\rm benv})^2$, as
compared with the case of solid rotation throughout the star.
This simple rotation model breaks down when it predicts ${\rm Co} \gtrsim 1$ at the base of the convective envelope,
and we employ the modified rotation profiles (\ref{eq:ominner}), (\ref{eq:ominnerb}).  


Considering first the $1\, M_\odot$ model, we compare several scenarios.  The star is either assumed to evolve in isolation;
or to interact with a Jupiter-mass planet that starts in a circular orbit of semi-major axis 1 AU.  The internal rotation
profile is either described by a two-layer model (Section \ref{s:RG_rotation}) during its post-MS evolution (`$j=c$' in the outer
envelope, with various inner $\Omega(r)$ profiles); or it is assumed to be solid-body from the ZAMS to the terminal AGB (`$\Omega=c$').


Figure \ref{fig:J_star_comparison_RGB_AGB} shows the evolution of $J_\star$.   We see that solid-body rotation
leads to a significant loss of angular momentum through the RGB winds, because little angular momentum is stored deep in the star.
Introducing an outer zone of uniform specific angular momentum allows only ${\rm 1/2}$ of the
angular momentum present at the end of the MS to be lost, with the remainder retained at the onset of core He burning.  The engulfment 
of the Jupiter greatly augments $J_\star$, but if $\Omega$ is uniform then most of this additional angular momentum is lost to RGB winds.

The most important consideration for us is whether ${\rm Co}_{\rm benv}$ reaches the threshold for dynamo activity.  The time
dependence of ${\rm Co}_{\rm benv}$ is shown in Figure \ref{fig:Co_base_comparison}. Only in the case of planet absorption with `$j=c$'
in the outer envelope does 
the star comfortably pass this threshold.  Enough angular momentum is absorbed from a Jupiter-mass planet that almost any profile
$\Omega(r) \propto r^{-\alpha}$ with $\alpha > 0$ will allow the star to attain ${\rm Co}_{\rm benv} = 1$ on the RGB.
Stronger angular momentum pumping is required near the tip of the AGB to sustain rapid
rotation: uniform rotation combined with absorption of a Jupiter does not suffice.

An isolated $M_{\rm ZAMS} = 1\,M_\odot$ star without a planetary companion does not attain ${\rm Co}_{\rm benv} \sim 1$ in its
inner envelope near the tip of the RGB or AGB, no matter the rotation profile.  This dynamo threshold is missed by more than two orders 
of magnitude in the case of solid rotation.

Angular momentum could also be transferred to an evolved giant by a tidal interaction with a stellar companion.  This happens
inevitably during the evolution to a common envelope and the formation of pre-CV system. For example, the $5M_\odot$ star may easily engulf a 
companion during the AGB phase, given the large maximum radius reached by the star ($\sim 600R_\odot$). We consider the
effect of injecting a companion of mass (3,10,30,100)$M_{\rm Jupiter}$ from an orbit with initial semi-major axis 
$a_i=2$AU when the star reaches a radius $R_{\star}=200R_\odot$ on the AGB.  
We also consider a range of inner rotation profiles, $\alpha = 1$ to $1.5$ in Equation (\ref{eq:ominner}).

Figure \ref{fig:R_c_and_Co_benv_for_M_companion} shows that the added angular momentum pushes $R_c$ outward, as expected; 
but when $\alpha = 1$ the relative flatness of the $\tau_{\rm con}(r)$ profile ($\tau_{\rm con}$ increasing roughly in proportion to $r$) 
may actually cause $\rm Co_{benv}$ to decrease.  In other words, the engulfment of a more massive 
companion may result in slower rotation at the base of the convective envelope.
Increasing the rotation index to $\alpha=1.5$ allows $\rm Co_{benv}$ to grow with increasing injected angular momentum
(Figure \ref{fig:R_c_and_Co_benv_for_M_companion2}).  
Our inference that angular momentum pumping in the envelopes of subgiants can sustain ${\rm Co}_{\rm benv} \sim 10$ 
(see Figure \ref{fig:rotation_profiles}) suggests that the rotation profile in the inner envelope may be significantly
steeper than $\alpha=1$ near the tip of the RGB and AGB.  This would imply an even stronger magnetization of the WD remnant, as we show in Paper II.

\section{Orbital Evolution of a Planetary \\ Companion to a $1M_{\odot}$ Star} \label{s:orbital_evolution}


Our calculation of the interaction of a planet with an evolving giant star follows the orbit in some detail.  We take 
into account the dissipation of the tide raised on the star, and gravitational scattering off a fluctuating quadrupole moment.  
The second effect, which has been previously considered in the context of massive neutron star binaries \citep{phinney92}, has a 
weaker radial dependence than the tidal torque and therefore modestly increases the range of semi-major axis over which a 
strong interaction between planet and star can take place.

We evolve the planet's orbit in response to these external forces, considering a range of masses (Earth, Neptune or Jupiter) 
along with different initial semi-major axes ($a_i = 0.5,0.75,1,1.5,2$ AU) and eccentricities ($e_i = 0,0.1,0.5$).  
The change in orbital angular momentum is deposited in the star and tracked self-consistently, including the effect of stellar mass
loss described by Equation (\ref{eq:djdtml}).




The tidal acceleration is calculated using \citep{hut1981}
\be\label{eq:atide}
{d^2{\bf r}\over dt^2} = -3k_2 \frac{G M_p}{r^2} \left(\frac{R_{\star}}{r}\right)^5 \bigg[ (1 + 3\frac{\dot{r}}{r}\tau)\hat{r} - 
 (\Omega_{\rm env} - \omega) \tau \hat{\phi} \bigg].
\ee
In the above expression, the position of the planet is followed in spherical coordinates, with $\omega$ its orbital angular
velocity and $M_p$ its mass.  We set the rotation frequency $\Omega_{\rm env}$ of the stellar envelope\footnote{Equation (\ref{eq:atide}) is 
derived assuming solid rotation in the envelope.} to zero.  The time lag of the tidal bulge is given by
\be
\tau \sim 2\frac{R_{\star}^3}{GM_{\star} \tau_{\rm con}} 
\ee
where 
\be \label{eq:convective_overturn_time}
\tau_{\rm con} \sim \left(\frac{M_{\rm env} R_{\star}^2}{L_{\star}}\right)^{1/3}
\ee
is the global convective overturn time.  The Love number $k_2$ is obtained (see Equations (11) of \citealt{zahn1989} 
and (17) of \citealt{scha1982}), 
\be
k_2 = 20.5 \alpha^{4/3} \frac{R_{\star}^{1/3} \tau_{\rm con}}{M_{\rm env}} \int_{x_{\rm benv}}^1 x^{22/3} L_r^{1/3} \rho^{2/3} l_P dx.
\ee
Here $\alpha = 2$ is the convective mixing length parameter, $M_{\rm env} = M_\star - M_{\rm benv}$ is the mass of the convective envelope of 
the star, $x = r/R_{\star}$, $x_{\rm benv} = R_{\rm benv} /R_{\star}$, and $L_r$ is the luminosity at a given radius.

The acceleration from the convective quadrupole is calculated using the standard expansion of the 
gravitational potential (see \citealt{murrd1999}),
\ba\label{eq:vquad}
V_{\rm quad} &=&  -\frac{G}{2r^5}\Bigl[(I_{yy}+I_{zz}-2I_{xx})x^2 \\ \nonumber
         &+& (I_{xx}+I_{zz}-2I_{yy})y^2 + (I_{xx}+I_{yy}-2I_{zz})z^2  \\ \nonumber
         &+&  6(I_{yz}\,yz + I_{xz}\,xz + I_{xy}\,xy)\Bigr].
\ea
(In practice we evolve the orbit of the planet in Cartesian coordinates.)  The moment of inertia tensor $I_{ij}$ 
is normalized using the rms convective quadrupole 
\be\label{eq:Q_conv}
\frac{Q_{\rm con}}{M_{\rm env} R_{\star}^2} \;\simeq\; 7.41 \times 10^{-3} f {L_{\star\,37}^{2/3}R_{\star\,13}^{5/3}\over
	(M_\star/M_\odot)(M_{\rm env}/M_{\odot})^{2/3}}.
\ee
Here $f$ is a factor of order unity which depends on the stellar structure.  This expression is obtained by integrating the density 
perturbation caused by convection inside the star over the envelope;  see the Appendix \ref{app:Q_conv} for a derivation.

\begin{figure}[!] 
\epsscale{1.0}
\plotone{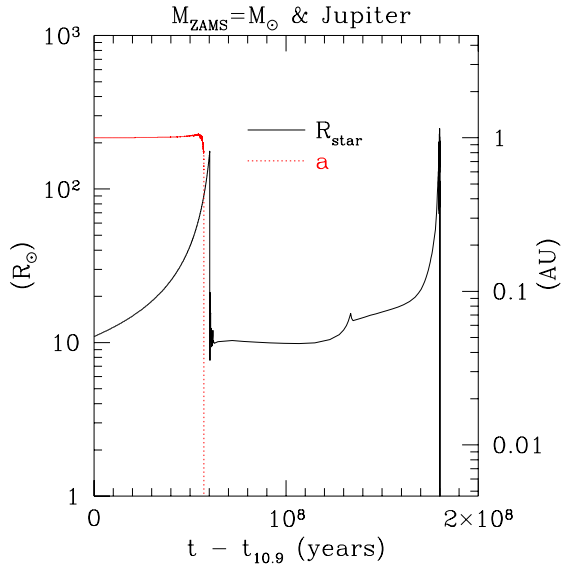}
\caption{Red line: orbital semi-major axis of a Jupiter-mass planet interacting with a RGB star and 
starting from a circular orbit with $a_i = 1$ AU.  Black line:  stellar radius.  Spiral-in of planet just 
below the tip of the RGB deposits angular momentum and triggers a dynamo process near the core-envelope boundary
(Paper II).}  
\vskip .2in
\label{fig:R_star_and_a_RGB_AGB}
\end{figure}

At time intervals separated by $\tau_{\rm con}$, the diagonal components of the moment of inertia matrix for the quadrupole were chosen randomly, 
with a magnitude uniformly distributed in the range [0, $Q_{\rm con}$]. The remaining components were then obtained by a 
random rotation of the diagonal matrix.  Given the random nature of this forcing effect, we ran ten realizations for each set of 
initial conditions to obtain a statistical sample of results.

A planet that is exposed to the tidal acceleration (\ref{eq:atide}) and quadrupolar potential (\ref{eq:vquad}) will evolve
in semi-major axis and eccentricity.  It can either collide with the envelope of the star or be ejected from the system. 
Here we are interested primarily in initial conditions leading to a collision.
In such an instance, we assume that drag and dynamical 
friction act instantaneously, resulting in the planet's engulfment and the transfer of its orbital angular momentum to the giant.

\subsection{Jupiter-mass Planet with $a_i = 1$ {\rm AU}}

Consider a Jupiter-mass planet starting at $a_i = 1$AU and $e_i = 0$. In Figure \ref{fig:R_star_and_a_RGB_AGB} we show one realization of its 
orbital evolution, comparing semi-major axis $a$ with $R_\star$.  In this case the planet is engulfed shortly before the star reaches the tip of
the RGB. We treat this as our fiducial model for a $M_{\rm ZAMS} = 1\,M_\odot$ star interacting with a planet.

\begin{figure}[!] 
\epsscale{1.0}
\plotone{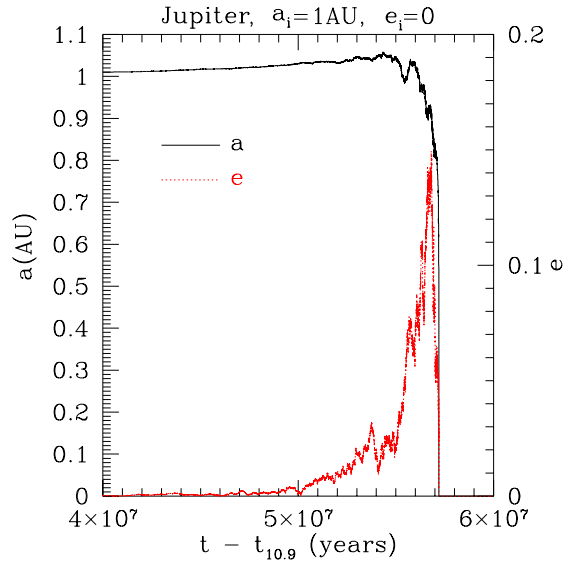}
\caption{Expanded view of the orbital parameters of the Jupiter-mass planet
in the same realization shown in Figure \ref{fig:R_star_and_a_RGB_AGB}.
Engulfment of the planet is preceded by a stochastic growth in orbital
eccentricity, due to gravitational scattering off the convective quadrupole.}
\vskip .2in
\label{fig:a_and_e_Jupiter_a0=1AU_e0=0}
\end{figure}

The corresponding eccentricity evolution is shown in Figure \ref{fig:a_and_e_Jupiter_a0=1AU_e0=0}. The stochastic nature of the convective
forcing of orbital parameters is evident here.

\subsection{Other Planetary Configurations}

\begin{figure}[!] 
\epsscale{1.0}
\plotone{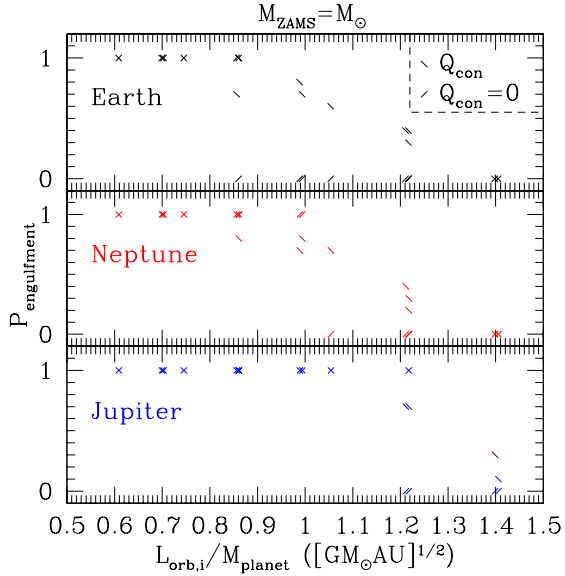}
\caption{Probability of engulfment of planets of various masses (Earth, Neptune, Jupiter), as a function of the initial specific orbital 
angular momentum.
Negative-sloping dash: planet scatters gravitationally off convective quadrupole of giant star.  Positive-sloping dash: $Q_{\rm con}=0$.
Probability of engulfment is enhanced by interaction with the quadrupole, especially for lower-mass planets.}
\vskip .2in
\label{fig:P_engulf_L_orb_0_and_non_0_Q_conv}
\end{figure}

Figure \ref{fig:P_engulf_L_orb_0_and_non_0_Q_conv} shows the probability of a planet's engulfment as a function of the initial 
(specific) orbital angular momentum.  Results for the three planetary masses we consider are shown separately.  The separate effect
of shutting off the convective quadrupole ($Q_{\rm con} \rightarrow 0$) is also shown.  

The convective forcing has an especially
strong effect on the probability of engulfment for Earth-mass planets.  In the case of a Jupiter-mass planet, there is a $\sim 20\%$
enhancement in the range of semi-major axes that results in a merger. These effects arise mainly from the excitation of eccentricity
and a corresponding reduction in periastron.

\section{Conclusions} \label{s:conclusions}

We have considered the inward pumping of angular momentum in a deep stellar convective envelope, with a 
goal of defining a workable rotation model for giant stars.
Independent theoretical and observational lines of argument point to a profile that differs radically from 
those obtained from stellar evolution codes such as MESA, in which the implementation of angular momentum transport
typically results in solid rotation within convective layers.  The work presented 
here and in Paper II also demonstrates
that much improved dynamo models must be implemented in order to capture basic features of stellar rotation, such as the relaxation to solid 
rotation in radiative parts of a star, and the rotational coupling between a radiative core and the base of an extended convective envelope.

The rotation of evolving stars cannot be understood even in qualitative terms without including the magnetic field built up
by a hydromagnetic dynamo operating near the boundary between convective and radiative zones.  Some features of this dynamo
are special to giant stars.  First, the high radiative energy flux leads to a strong broadening
of the tachocline layer situated beneath the core-envelope boundary, which may facilitate mixing of core material
into the envelope. The growth of the core is also accompanied by an inward flux of magnetic helicity. These effects are investigated in 
detail in Paper II, where we motivate a strong magnetic coupling between core and inner envelope when ${\rm Co}_{\rm benv} \gtrsim 0.1-0.3$.  

When this condition is satisfied, a simple picture emerges:  nearly solid rotation in the radiative parts of the 
star that previously supported a dynamo, combined with an inhomogeneous rotation profile in a deep convective envelope,
whose slope depends on the local Coriolis parameter.  For example, solid rotation in a radiative layer adjacent to
convective zone is facilitated by a tiny helical magnetic field.  Differential rotation that is sourced
by a changing stellar mass profile during subgiant expansion can be significantly reduced if the poloidal magnetic 
pressure exceeds $\sim 10^{-16}$-$10^{-15}$ of the convective pressure, increasing to $\sim 10^{-8}$
during the thermally pulsating AGB.  Magnetic fields $\sim 10^2$-$10^3$ times stronger are required to
compensate a latitude-dependent convective stress that is partly transmitted through a tachocline layer (Paper II).

A magnetic field exceeding this minimal value is plausibly present in the radiative core of a subgiant:
erasure to a lower level would require a very efficient mechanism for eliminating magnetic helicity.  Inhomogeneous
rotation could, in principle, be maintained by a supplementary mechanism of angular momentum transport such as 
gravity wave damping.  

Our results can be summarized as follows.

1.   Strong inward pumping of angular momentum is well motivated by measurements of core rotation in stars of radius 
$\lesssim 10\,R_\odot$ \citep{becketal2012,mossetal2012}, and by numerical simulation \citep{brunp2009}.  A rotational decoupling between 
core and envelope (as envisaged e.g. by \citealt{CantMBCP2014}) is difficult to achieve over the long duration of subgiant expansion,
depending on an unrealistically small poloidal magnetic field.  We have found that a profile $\Omega(r) \propto r^{-1}$ within parts of the 
envelope that are very rapidly rotating (${\rm Co} \gtrsim 1$), combined with solid rotation in the radiative core, provides an excellent fit to the data for stars more massive than $\sim 1.3\,M_\odot$.  This corresponds to the profile
that is derived from quasi-geostrophic balance in an adiabatic envelope with heat transported by radially
extended convective upflows and downflows that do not mix significantly with each other over a scale height.
Given a gravity profile $g(r) \propto r^{-\beta}$ with $\beta \simeq 1$, we deduce $\Omega(r) \propto r^{-(1+\beta)/2}
= r^{-1}$.

More focused measurements of individual stars can in principle be used to test this rotation model -- in particular, to 
constrain the range of $\Omega$ within the convective envelope -- and also to test for the existence of strong
differential rotation across the core-envelope boundary, as in the model of \cite{CantMBCP2014}.  Efforts so far to
measure the full radial profile $\bar\Omega(r)$ have considered separate zones of solid rotation in core and envelope, as
well as smoother profiles; the fits are mainly sensitive to the rotation rate in the core \citep{deheuvels12,deheuvels14}.

Solar-mass stars which spin down
significantly on the MS, and do not absorb a planetary compansion, are predicted to retain much slower core rotation
during subgiant expansion, but to rotate within the range of rates inferred from {\it Kepler} observations of
core He burning stars. Although the {\it Kepler} subgiant sample is mainly composed of stars more massive than $1.3\,M_\odot$, this provides a sharp test of the rotation model presented here.  On the other hand, we note that convergence in the core rotation of stars below and above the $1.3\,M_\odot$ threshold during subgiant expansion would be remarkable in almost any rotation model, given the significant
difference in rotation rate at the end of the MS.

2.  Independent constraints on internal rotation in subgiants come from measurements of the slowdown in surface
rotation as stars of mass $\gtrsim 1.3\,M_\odot$ first ascend the giant branch.  A good part of the discrepency,
found by \cite{schrijver93}, with stellar models which assume solid rotation, is removed by introducing the
same envelope rotation profile that is implied by the {\it Kepler}
asteroseismic data.   A separate rapidly rotating core, combined with solid rotation in the envelope, has a negligible
effect on surface rotation because $I_{\rm core}$ is a very small fraction of $M_{\rm benv} R_{\rm benv}^2$ (and of the effective moment of 
inertia of the envelope).

3.  Stars near the tip of the RGB and AGB (radii $\gtrsim  10^2\,R_\odot$) probe a different regime of slow rotation that
is not directly constrained by the {\it Kepler} observations.  Here we have considered steeper rotation profiles,
$j \sim$ const, in the outer parts of the envelope where ${\rm Co} \lesssim 1$.  This stronger level of angular momentum pumping
allows rapid rotation to be sustained deep in the envelope even during the final stages of envelope ejection.  It also limits the loss of 
angular momentum to winds on the RGB and AGB.  We find that the conditions
for a hydromagnetic dynamo  are favorable in a star which rotates rapidly at the end of the MS (e.g. initial mass
$\gtrsim 1.3\,M_\odot$); or in a solar-mass star which absorbs a Jupiter (or even Neptune) at some point
during its post-MS expansion.

The gravity profile steepens significantly in the inner envelope following the first dredge-up phase, and approaches
$g(r) \propto r^{-2}$ near the tips of the RGB and AGB.  If the convective upflows and downflows remain radially extended,
then our scaling solution to the vorticity equation implies that the inner angular velocity profile steepens to 
$\Omega(r) \propto r^{-3/2}$.

4. The mass of an individual convective cell in a giant envelope greatly exceeds the mass of Jupiter. This means than the
orbital eccentricity of any planet will be significantly excited by the associated gravitational quadrupole.  We have
calculated the orbital evolution of a suit of planets of various masses (Earth, Neptune, Jupiter) in response to this
gravitational stirring as well as tidal drag.  The gravitational stirring has a significant effect on the probability
of absorption of an Earth-mass planet starting at a semi-major axis 1 AU.  

5. The case of a solar-mass star which spins down on the MS and then {\it does not} regain angular momentum
from a companion is more complicated.  The rotation model described here can be applied while the star
remains more compact than $R_\star \sim 10^{12}$ cm; but slow rotation is enountered throughout the convective
envelope near the tip of the giant branches.  In these most extended phases of expansion,
the coupling between core and envelope will depend on additional (and perhaps largely hydrodynamic) mixing mechanisms, whose exact nature is 
still being debated \citep{maeder00,fuller14}.

More extended simulations of deep convection, including radiative transport and feedback from 
magnetic fields, are required to confirm the dependence of
rotation profile on Coriolis parameter.  A more centrally peaked profile is plausibly
achieved near maximum expansion on the giant branches, where the effects of rotation are weaker, but
whether this profile really approaches uniform specific angular momentum needs further testing.
We emphasize the degeneracy between the combined effects of
flattening the slope of the rotation profile and increasing the spindown on the MS (both of which
reduce the rotation rate in the inner envelope), and raising the angular momentum that is received from
a sub-stellar companion during post-MS expansion.   Although magnetic field amplification in the inner envelope 
becomes more difficult as the rotation profile flattens, even a planetary companion can have a significant compensating effect.

\acknowledgements This work was supported by NSERC.  We thank the anonymous
referee for detailed comments on the presentation of our results.

\begin{appendix}

\section{Normalization of Convective Quadrupole During Giant Evolution} \label{app:Q_conv}

The internal density, temperature and luminosity profiles of a model giant star 
evolve with time in a complicated way.  Here we describe a simple prescription
for the net gravitational quadrupole that is induced by convective motions in the giant envelope, 
which can be expressed in terms of integral quanties such as stellar luminosity $L_\star$, radius $R_\star$ and core and envelope masses $M_{\rm core}$, $M_{\rm env}$.

We estimate the magnitude of the quadrupole from
\be\label{eq:qcon}
Q_{\rm con} \simeq \sqrt{N_{\rm cells}} \left<\delta M r^2\right>
\ee
where $N_{\rm cells}$ is the number of convective cells in the envelope of the star, and
$\delta M$ is the mass perturbation induced by convection in a cell, and the average is over the volume
of the envelope.  We assume that the convective cells (of angular size $\Delta\Omega$) penetrate the whole star, so that
\be
N_{\rm cells} \sim \frac{4\pi}{\Delta\Omega};\quad\quad
\left<\delta M r^2\right> \sim \Delta\Omega \int_0^{R_{\star}} r^4 \delta \rho dr.
\ee
The density perturbation $\delta \rho \sim \rho v_{\rm con}^2/c_s^2$, where we estimate $\rho v_{\rm con}^3 \simeq L_{\star}/4\pi r^2$.
Therefore
\be
\left<\delta M r^2\right> \sim \Delta \Omega \left( \frac{L_{\star}}{4\pi} \right)^{2/3} \int^{R_{\star}}_0 r^{8/3} \frac{\rho^{1/3}}{c_s^2}dr,
\ee
where we take $\Delta\Omega = 1$.  

The assumption of a point gravitational potential, in combination with an adiabatic equation of state, gives
a good fit to the actual density profile of a giant.  The temperature profile is more sensitive to the mass
distribution in the star.  Hence
\be
\rho(r) = \frac{4M_{\rm env}}{\pi^2 R_{\star}^3} \left( \frac{R_{\star}}{r} - 1 \right)^{3/2};\quad\quad
c_s^2(r) = \frac{2}{3} \frac{G}{R_{\star}} M(<r) \left( \frac{R_{\star}}{r} - 1 \right) = 
{2GM_\star\over 3R_\star} g\left({r\over R_\star},{M_{\rm core}\over M_\star}\right) \left({R_\star\over r}-1\right),
\ee
where $M(<r)$ is the enclosed mass, which is proportional to 
\be
g(x,M_{\rm core}/M_\star) = {M_{\rm core}\over M_\star} + {16M_{\rm env}\over \pi M_\star} \int_0^x {x^\prime}^2 
\left( {1\over x'} - 1 \right)^{3/2} dx^{\prime}.
\ee

The quadrupole (\ref{eq:qcon}) works out to
\be
\frac{Q_{\rm con}}{M_{\rm env} R_{\star}^2} \simeq \frac{3}{4^{1/3} \pi^{5/6} G} \frac{\Delta \Omega^{1/2} L_{\star}^{2/3} 
R_{\star}^{5/3}}{M_{\rm env}^{2/3} M_{\star}} f; \quad\quad
f \equiv \int_0^1 {x^{8/3}\over g(x,M_{\rm core}/M_\star)} \left( \frac{1}{x} - 1 \right)^{-1/2} dx.
\ee

\end{appendix}

\end{document}